# Matrix metalloproteinases as new targets in Alzheimer's disease: Opportunities and Challenges

*Pauline Zipfel,[†] Christophe Rochais,[†] Kévin Baranger,[‡] Santiago Rivera,[‡] Patrick Dallemagne[†*]*

† Centre d'Etudes et de Recherche sur le Médicament de Normandie (CERMN), Normandie Univ, UNICAEN, 14000 Caen, France.

‡ Aix-Marseille Univ, CNRS, INP, Inst Neurophysiopathol, Marseille, France.

**ABSTRACT**

Although matrix metalloproteinases (MMPs) are implicated in the regulation of numerous physiological processes, evidences of their pathological roles have also been obtained in the last decades, making MMPs attractive therapeutic targets for several diseases. Recent discoveries of their involvement in central nervous system (CNS) disorders, and in particular in Alzheimer's disease (AD), have paved the way to consider MMP modulators as promising therapeutic strategies. Over the past few decades, diverse approaches have been undertaken in the design of therapeutic agents targeting MMPs for various purposes, leading, more recently, to encouraging developments. In this article, we will present recent examples of inhibitors ranging from small molecules and peptidomimetics to biologics. We will also discuss the scientific knowledge that has led to the development of emerging tools and techniques to overcome the challenges of selective MMP inhibition.

## 1. Introduction: General remarks about Alzheimer's disease nowadays

Alzheimer's disease (AD), the most common form of dementia (~70%) in the elderly, is a chronic and neurodegenerative brain disorder characterized by memory loss and other cognitive impairments. In 2018, the number of people living with dementia in the world was estimated at 50 million and this number will likely more than triple to 152 million by 2050.[1] With only 4 drugs approved (Donepezil, Galantamine, Rivastigmine and Memantine) to barely relieve the symptoms of AD, finding a cure, or at least a treatment that delays the progression of the disease, remains a real challenge for the community. Because of the heavy economic and societal impacts, there is an urgent need to find new treatments that target the molecular causes of the neurodegenerative process. During the past three decades, scientists have debated about the true molecular causes of the disease, but these remain elusive to this day. Two abnormal aggregates of beta-amyloid peptide (Aβ) and hyperphosphorylated Tau protein, respectively, constitute what has been considered for many years the main pathological features of AD: extracellular amyloid plaques (also known as senile plaques) and intracellular neurofibrillary tangles. Other pathophysiological disturbances in AD have also been pointed out, such as alterations in cholinergic and glutamatergic neurotransmission, neuroinflammation, oxidative stress and mitochondrial dysfunctions. In 2020, it is still unclear how these interconnected events influence each other, why they become damaging over time or what is their exact chronology. Thus, ongoing experimental and clinical research continues to expand knowledge of potential new biological targets involved in the pathogenesis of AD.

In this context, matrix metalloproteinases (MMPs) have been recently highlighted in the literature as potential new relevant biological targets in AD. In this paper, we will review the pathophysiological activities of MMPs in the central nervous system (CNS) in general and more specifically in AD. We will then focus our attention on the challenges to be addressed when targeting MMPs and discuss the different strategies and their evolution for the design of therapeutic agents that modulate their activities.

## 2. Description of the matrix metalloproteinase family

MMPs, also known as matrixins, form a family of endopeptidases characterized by the presence of a zinc cation stabilized by interactions with three histidine residues in their catalytic site. They belong to the larger metzincin superfamily of metalloproteinases and are present in most tissues of the body. To date, there are more than twenty MMPs described in humans (Table 1). They are classified according to their abilities to cleave substrates (initially discovered) in collagenases, gelatinases and matrilysins, and according to the localization (stroma) where they were first identified for stromelysins. Most MMPs are secreted, but a group of six proteinases bound to the membrane are referred to as membrane-type MMPs (MT-MMPs). Finally, a more heterogeneous group is often classified as "others".

| Class | Numbering |
|---|---|
| **Collagenases** | MMP-1, MMP-8, MMP-13 |
| **Gelatinases** | MMP-2, MMP-9 |
| **Stromelysins** | MMP-3, MMP-10, MMP-11 |
| **Matrilysins** | MMP-7, MMP-26 |
| **MT-MMP (membrane type)** | MMP-14 (MT1-MMP), MMP-15 (MT2-MMP), MMP-16 (MT3-MMP), MMP-17 (MT4-MMP), MMP-24 (MT5-MMP), MMP-25 (MT6-MMP) |
| **Others** | MMP-12, MMP-19, MMP-20, MMP-21, MMP-23, MMP-27, MMP-28 |

**Table 1.** MMPs classification in humans, adapted from [2] and in accordance with UniProt database (www.uniprot.org).

All these MMPs are synthesized with a common N-terminal signal sequence, which is then cleaved in the endoplasmic reticulum to form the latent proenzymes. MMPs are made up of at least two domains, which are the pro-domain (containing the so-called "cysteine-switch" sequence motif, Pro-Arg-Cys-Gly-Xxx-Pro-Asp, except for MMP-23) and the zinc-containing catalytic domain. The cysteine residue of the pro-domain binds to the catalytic zinc as the fourth ligand in tetrahedral coordination sphere and thus keeps MMPs in their inactive form. Class-specific domains complete their structure (hemopexin-like domain, hinge region, collagen-binding domain, membrane anchored domain…).[3,4] These proMMPs, also called zymogens, are activated by different mechanisms and subsequently either secreted in the extracellular space or attached to the cell membrane, although new intracellular MMP activities have also been reported.[5] The activation step includes most of the time the "cysteine switch" event, consisting in successive cleavages inside the pro-domain by trypsin, plasmin or other MMPs, which results in the release of the pro-domain and the disruption of the Cys-Zinc interaction. Other activation mechanisms have been reported, including the formation of complexes between proMMPs and endogenous tissue inhibitors of MMPs (TIMPs), as well as furin cleavage in the trans-Golgi network.[3,6]

MMPs are involved in the proteolysis of extracellular matrix components, in addition to a large number of non-matrix substrates such as growth factors, cytokines, chemokines, cell surface proteins receptors – to name just a few – which may become activated or inactivated. Thus, depending on their localization and their substrate specificity, MMPs have physiological functions involved in homeostatic processes such as development, morphogenesis and tissue reorganization.[7,8] Like all proteinases, their activities must be tightly regulated because of their biological importance and potency. The activation/inhibition balance is maintained at several levels in vivo, including gene expression, proMMPs activation and endogenous inhibition in the extracellular medium by the family of TIMPs (TIMP-1 to TIMP-4), as well as by other inhibitors such as α2-macroglobulin, a plasma inhibitor.[7,9,10] The four TIMPs are macromolecules containing about 190 amino acids, which form equimolecular complexes with individual MMPs in their active form depending on their specificity and thereby block the active site of MMPs.[11]

In some situations, the disruption of the MMP/TIMP balance is observed, as upregulation of MMP activity has been associated to many disorders such as cancer, arthritis, scarring processes, atherosclerosis, infections, inflammatory and immune diseases,[12,13,14] for which the development of specific MMP inhibitors has been validated as a genuine therapeutic strategy (e.g., MMP-1, -2 and -7 in cancer).[15,16] More recently, MMP involvement has also been confirmed in neurodegenerative diseases.[10,17,18]

## 3. Involvement of MMPs in the central nervous system physiology

In the CNS, MMPs are expressed at a modest level in physiological conditions and their regional distribution varies depending on the MMP. They are produced by all brain cells, e.g., endothelial cells, oligodendrocytes, astrocytes, microglia and neurons[19] and contribute together with TIMPs to nervous system physiology during ontogenesis, neurogenesis, angiogenesis and neuronal plasticity.[5,20] MMPs implication in neuronal plasticity has been mainly linked to their ability to influence learning and memory and what is believed to be the underlying cellular substrate, long-term potentiation (LTP).[5,21,22] MMP-9, MMP-3 and MT5-MMP have been mainly studied in this context. Moreover, MMPs (e.g., MMP-2, MMP-9, MT3-MMP, MT5-MMP) have been also involved in synaptogenesis,[22,23,24] migration of neural cells and their precursors,[25,26,27] and nervous tissue regeneration.[26,28,29,30,31,32]

However, MMPs are ambivalent enzymes that can also exert negative effects depending on the biological context.[5] Along this line, upregulation of MMP expression has been extensively documented in a myriad of pathological processes including excitotoxic epileptic seizures,[22,33,34] hypoxia/ischemia,[35,36] neuronal death,[35,37] microbial infections,[38] blood-brain barrier (BBB) disruption,[35,38,39] neuroinflammation,[40,41] demyelination[5] as well as glioma progression.[42] MT1-MMP also seems to influence familial amyloidotic polyneuropathy, a rare, systemic disease with autosomal dominant transmission, due to a mutation in the transthyretin gene.[43] Various MMPs are also involved in several chronic neurodegenerative diseases, such as Parkinson's disease, amyotrophic lateral sclerosis, Huntington's disease, multiple sclerosis and AD.[17,18,44] Neuroinflammation is systematically a common feature of these various brain pathologies, during which the production and activation of specific MMPs are initiated or amplified by neural

(astrocytes, microglia, endothelial cells…) or immune cells (macrophages, lymphocytes, neutrophils…).[40]

Because MMPs are multifaceted enzymes that mediate a myriad of physiological and pathological pathways, the risk/benefit outcome must be carefully estimated when considering their inhibition in order to prevent/anticipate unwanted side effects. Better understanding the molecular interactions of MMPs in a given spatio-temporal setting is therefore a pre-requisite to implement innovative drug discovery strategies that interfere with their harmful effects, while sparing their physiological actions. These topics have been discussed in detail for a number of CNS disorders and physiological conditions in a series of recent reviews.[18,38,39,41,45]

## 4. MMPs in Alzheimer's disease

There is growing evidence that several MMPs may contribute to or interfere with the pathophysiological mechanisms of AD. The following paragraphs will provide an update on the latest developments in this area, based on knowledge gained directly from AD patients or in vitro/in vivo models of AD.

### General implication of MMPs in AD

Elevated expression of different MMPs has been reported in AD brains, including MMP-1, MMP-2, MMP-3, MMP-9, MMP-13 and MT1-MMP (and also MT5-MMP that we will address separately).[46,47,48,49,50] The effects of these MMPs have been mostly related to their functional interactions with Aβ or Tau. Aβ results from the proteolytic processing of amyloid precursor protein (APP), a type I transmembrane protein of 695 to 770 amino acids, whose physiological functions are not yet fully elucidated.[51,52] There are two main physiological proteolytic pathways for APP that are mediated by proteases called secretases: the non-amyloidogenic pathway, which is neuroprotective, precludes Aβ production and is predominant under physiological conditions, and the amyloidogenic pathway, which leads to the production and eventually neurotoxic accumulation of the Aβ peptide under pathological conditions (Figure 1).[53]

In the non-amyloidogenic pathway, APP is first cleaved at the α-site by an α-secretase (mainly ADAM-10 and -17, for a disintegrin and metalloprotease domain), leading to the extracellular release of a soluble N-terminal fragment sAPPα with neurotrophic properties, and a residual membrane-associated C-terminal fragment α (CTFα, also called C83). Cleavage of C83 within the transmembrane domain by the γ-secretase complex (consisting of presenilin 1 or 2, nicastrin, anterior pharynx-defective-1 and presenilin enhancer-2) generates an extracellular non-toxic p3 fragment and an APP intracellular domain (AICD) fragment that is unstable and rapidly degraded in this processing pathway.[54]

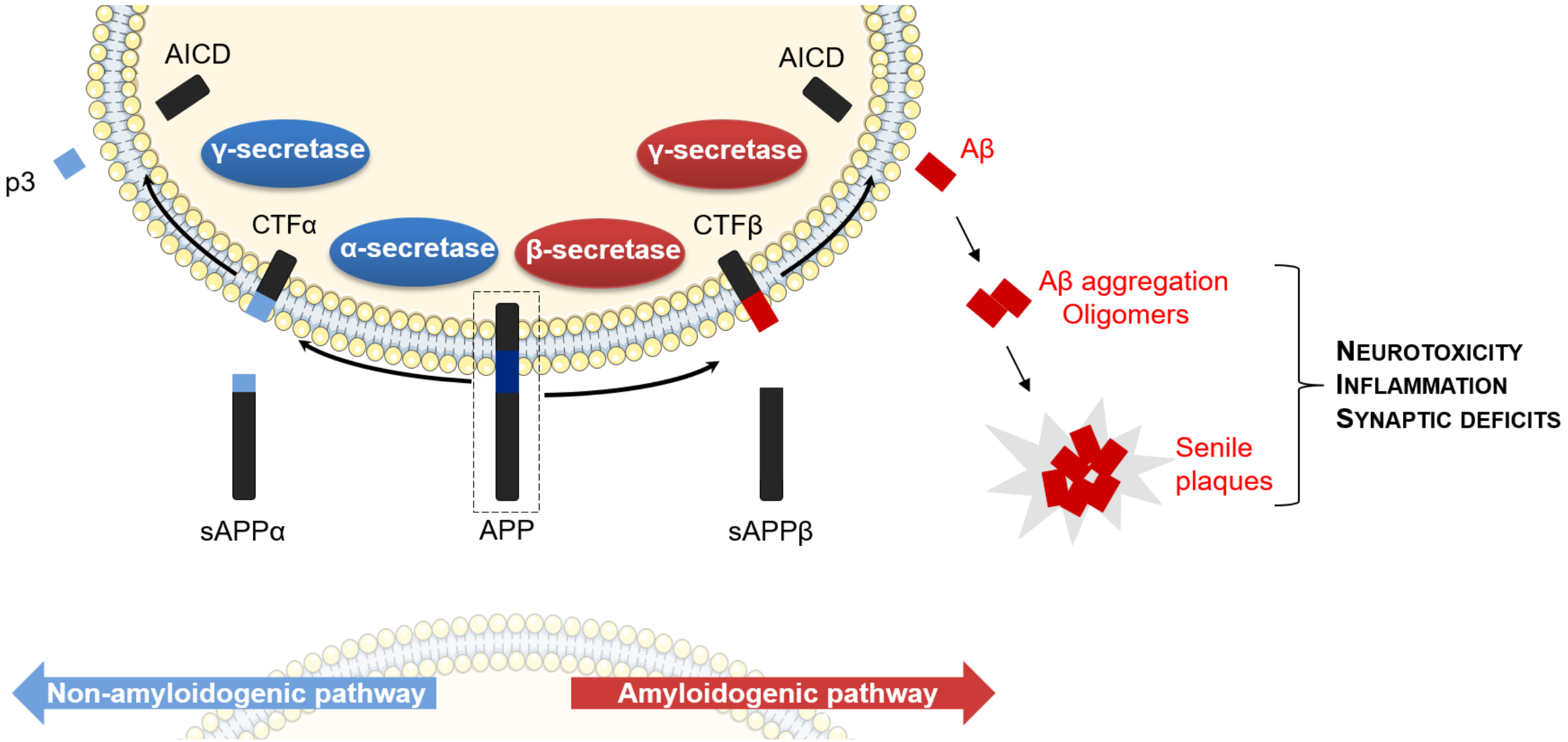

**Figure 1.** Schematic representation of the two classical APP processing pathways. In blue, the non-amyloidogenic pathway, which mainly occurs at the plasma membrane level and in red, the amyloidogenic pathway, which mainly occurs in the endosomes.

APP can also undergo amyloidogenic processing when cleaved at its β site by a β-secretase (mainly BACE-1 for beta-site APP cleaving enzyme 1), thus generating the secreted ectodomain sAPPβ and the neurotoxic membrane-associated CTFβ fragment (also called C99).[55] Subsequent cleavage of CTFβ by the γ-secretase results in the formation of Aβ and the AICD fragment. The latter can be translocated to the nucleus, where it could be involved in the regulation of gene expression and other processes, e.g. apoptosis.[54,56] The levels of Aβ are tightly regulated by two main degradation proteolytic pathways that take place extracellularly or intracellularly after receptor-mediated endocytosis. MMPs are among the Aβ-degrading enzymes, along with neprilysin, insulin-degrading enzyme, angiotensin-converting enzyme and endothelin-converting enzyme.[57] In pathological conditions such as AD, different pathogenic events result in an imbalance between Aβ production and clearance. Thus, abnormal accumulation of Aβ in the form of oligomers or amyloid plaques is believed to play a central role in the pathogenesis of AD and to be responsible for synaptic and cognitive deficits as well as neuroinflammation.[58]

Among the Aβ-degrading enzymes, MMP-2 and MMP-9 expressed in reactive astrocytes surrounding amyloid plaques, have demonstrated their in vitro and in vivo ability to degrade soluble Aβ and amyloid plaques into non-toxic fragments (Table 2), possibly illustrating a mode of action by which these MMPs could help preserve neurons from amyloid toxicity under AD conditions.[47,59,60,61,62] This is also partly inferred from in vivo studies where genetic deletion (knock-out mice) or pharmacological inactivation of MMP-2 and MMP-9 in non-AD mice caused increased levels of Aβ.[63] Clearly, analogous MMP-2/MMP-9 knockout or knockdown experiments are still lacking in transgenic AD mouse models. Meanwhile, reduced amyloid pathology following immune-based therapy in an AD mouse model has been associated to an intriguing increase of MMP-9 levels in the brain.[64] In pace with a potential beneficial effect of MMP-9 in AD, it has been shown that its overexpression in a mouse model of AD promotes the increase in the levels of neuroprotective sAPPα, along with a decrease in Aβ oligomers and the improvement of cognitive abilities.[65] This study is interesting from a biochemical standpoint because it reflects the possibility that APP is a physiological substrate of MMP-9, which could occasionally behave as a α-secretase-like enzyme. However, chronic overexpression of MMP-9 as a valuable therapeutic approach is arguable, since MMP-9, unlike MMP-2, has been found to be neurotoxic for hippocampal neurons.[34] Moreover, an increase in cortical MMP-9 activity during early AD has been shown to correlate with cognitive deficits, possibly related to proteolytic degradation of nerve growth factor.[66] Further caution is advised by data indicating that detrimental brain hemorrhages following anti-Aβ immunotherapy trials are associated with the upregulation of MMP-9 levels and activities,[67] which is consistent with the pro-inflammatory role of MMP-9 and its contribution to BBB breakdown.[39]

In line with this idea, the presence of the ε4 allele of apolipoprotein E (ApoE) in the genome – a major genetic risk factor for AD – leads to BBB breakdown through the proinflammatory cyclophilin A pathway that involves the activation of MMP-9.[68] The upregulation of MMP-2 activity has also been linked with BBB leakage[39] and it is known that oligomeric Aβ can stimulate the levels of MMP-2 in astrocytes surrounding amyloid plaques, certainly mediated by the release of proinflammatory cytokines by microglia.[69,70] Altogether, these data highlight the possibility that both MMP-2 and MMP-9 could exert detrimental effects in AD through the induction of BBB leakage, which may in turn support chronic inflammation. BBB breakdown has been controversial for years in the AD field due to the absence of massive leukocyte infiltration as it is observed in other disorders such as multiple sclerosis, stroke or epilepsy.[71,72] However, recent works reporting BBB dysfunctions in animal models of AD[73] and in Humans[72], point out the necessity of considering BBB demising factors as potential therapeutic targets.

In the series of other possible adverse effects of MMP-2 and MMP-9 in AD, the upregulation of MMP-2, at early and middle stages of AD in neurons containing neurofibrillary tangles, has raised the suspicion that the proteinase could promote the pathogenic accumulation of Tau aggregates.[74] Similar hypotheses have been put forward in brains from AD patients, where upregulated MMP-9 expression has been found associated with neurofibrillary tangles.[75]

It is reasonable to suggest that sustained inhibition of MMP-2, which is constitutively expressed in the brain, may be consistent with the chronic nature of AD progression, whereas transient upregulation of MMP-9 could be better targeted in a narrower time frame, for example in the context of peripheral infection leading to accelerated cognitive decline in AD patients, which is likely related to the opening of the BBB and concomitant neuroinflammation.[76] In any case, the development of specific inhibitors for these MMPs could help clarify their function in AD, taking into account their spatio-temporal pattern of expression/activity, which has yet to be mapped in detail.

MT1-MMP expression is strongly upregulated in the brain of AD mice in reactive astrocytes surrounding amyloid deposit,[69] as well as in microglia/macrophages[50] and neurons[47]. Like MMP-2 and MMP-9, MT1-MMP seems to display a dual functionality in AD. Thus, exogenously added to cell cultures, MT1-MMP can degrade soluble and aggregated Aβ species in vitro and in situ,[69] but if overexpressed in HEK cells carrying the APP Swedish mutation, the enzyme leads to an increase in CTFβ and Aβ levels, thereby highlighting its potential contribution to AD pathogenesis.[47] The pro-amyloidogenic effect of MT1-MMP involves a β-secretase-dependent mechanism and/or the promotion of APP trafficking into endosomes, where Aβ production mainly occurs.[77] In vivo confirmation of the functional benefits of MT1-MMP inhibition is necessary and experimentally within reach. For example, through genetic or pharmacological approaches that suppress or inhibit MT1-MMP activity. In the former case, since MT1-MMP deletion causes mice lethality shortly after birth,[78,79] conditional MT1-MMP deficient mice should therefore be generated in an AD background. In the second case, selective MT1-MMP blocking antibodies[80,81] could be infused in AD mice at prodromal-like stages of the pathology.

A series of studies reported that APP contains a proteinase inhibitor domain for MMP-2 located in the C-terminal glycosylated region of soluble APP forms.[82] Inside this domain, a decapeptide sequence Ile-Ser-Tyr-Gly-Asn-Asp-Ala-Leu-Met-Pro termed APP-derived peptide inhibitor (APP-IP) specifically inhibits MMP-2.[83,84,85] Interestingly, the conversion of pro-MMP-2 into the active MMP-2 form was prevented by a fusion protein containing the APP-IP in TIMP-2.[86] The chimera combined potent APP-IP-mediated inhibition of the MMP-2 catalytic site and selective recognition of the MMP-2 hemopexin domain by TIMP-2. It is noteworthy that not only can MT1-MMP cleave APP upstream of the APP-IP to release soluble APP lacking APP-IP,[77,87] but this process can be performed in cooperation with MMP-2.[77] This novel functional interaction between both MMPs adds to the well-known ability of MT1-MMP to catalyse the conversion of inactive pro-MMP-2 to its active MMP-2 form.[88] In this context, an increase in active levels of MMP-2 mediated by MT1-MMP could promote the formation of sAPP without APP-IP, resulting in a reduced ability of the system to inhibit MMP-2 that could further contribute to the cleavage of APP and so on. Functional links between MT1-MMP and MMP-2 could broaden the spectrum and potency of their pathogenic actions, but also expand the possibilities of specifically targeting either proteinase to interfere with the proteolytic cascade.

If MMP-2, MMP-9 and MT1-MMP act as Aβ-degrading enzymes as well as pathogenic effectors, is there any justification for activating or inhibiting these proteases? Promoting MMP activity, especially chronically, may be inherently risky. Indeed, the natural inhibitors, TIMPs, usually exceed MMPs content. In addition, MMPs are mainly present in their inactive forms. All this together implies that keeping a tight control on MMP activity appears to be essential to preserve tissue homeostasis. A hypothetical degradation strategy of Aβ based on the activation of MMPs could result in the indiscriminate degradation of many physiologically important substrates, in addition to the activation/amplification of difficult to control proteolytic cascades involving MMPs or other proteinases (e.g., serine proteinases), characteristic of many pathogenic processes. Therefore, it is likely that promoting MMP activity may eventually increase the risk/benefit ratio. On the other hand, inhibiting a given MMP could also affect a number of physiologically relevant off-target substrates and have side effects. This is for instance one of the main obstacles to AD therapies aiming at inhibiting γ-secretase to block the generation of Aβ (Figure 1), because the catalysis of other substrates by this enzyme supports physiological functions.[89] In the case of MMPs inhibition, one would expect that the biochemical/biological redundancy of this family of proteinases would limit these drawbacks, as other MMPs could take over the cleavage of a physiological substrate affected by the specific inhibition of a particular MMP. Overall, specific inhibition of MMPs appears to be a more appropriate strategy than their activation or upregulation for the prospects of potential therapeutic interventions.

The possible implication of MMP-13 in AD has been recently highlighted,[90] as elevated levels of the enzyme were found in the brain of AD patients and AD mice.[49] The same work linked the involvement of MMP-13 in amyloid pathology by its ability to regulate BACE-1 expression through the phosphatidylinositol kinase-3 (PI3K) signaling pathway. In addition, downregulating MMP-13 activity through pharmacological inhibition or MMP-13 knockdown in AD mice improved amyloid pathology and cognitive deficits.[49] Previously, another study showed that Aβ peptide could induce MMP-13 in microglial cells through a PI3K/Akt-dependent mechanism,[91] likely implying a regulatory feedback between MMP-13 and Aβ.

Other MMPs have also proven to respond to Aβ. Thus, cultured microglia from post-mortem AD brains exposed to Aβ showed increased levels of MMP-1, MMP-3, MMP-9, MMP-10 and MMP-12, possibly indicating the involvement of these MMPs in AD associated neuroinflammation,[92] but also in neurodegeneration as suggested for MMP-1.[93]

Despite a growing number of reports linking elevated levels of MMPs in different CNS areas with the progression of AD,[18] there is a clear need for detailed mapping of the regulation of individual MMPs at different stages of disease. Taken together, these data should consolidate expectations about the importance of MMPs in AD and prefigure functional specificities of each proteinase in the disease context.

**Table 2.** Overview of the main pathophysiological functions of MMPs in AD and in other pathological and physiological settings.

| MMP | Producing cells implicated in AD | Pathophysiological roles in AD | Other functions |
|---|---|---|---|
| **MMP-1** | Neurons; Astrocytes; Microglial cells; Endothelilal cells; Immune cells | Neuroinflammation; neuronal death[93] | Cancer[2,15,16]; Atherosclerosis[94,95]; Immunity[96]; Vascular remodeling[97] |
| **MMP-2** | Astrocytes; Neurons; Microglial cells; Endothelial cells; Oligodendrocytes; Immune cells; | Aβ degradation[47,62,63]; Neuroinflammation[69,70]; Associated with Tau aggregates[74,18]; Associated with Aβ deposits[47] | Cancer[2,15,16]; Learning and Memory disabilities[98]; Neuroinflammation[99]; Atherosclerosis[94,95]; Vascular remodeling[97] |
| **MMP-3** | Astrocytes; Neurons; Microglial cells; Endothelial cells; Oligodendrocytes; Immune cells | Aβ degradation[100]; Blood-Cerebrospinal fluid barrier degradation[101,18] | Cancer[2,15,16]; Synaptogenesis[102]; α-synuclein cleavage[103]; Neuroinflammation[104]; Atherosclerosis[94,95]; Vascular remodeling[97] |
| **MMP-9** | Astrocytes; Neurons; Microglial cells; Endothelial cells; Oligodendrocytes; Immune cells | Vasculature damages[68]; Aβ degradation[60,61,62,63]; sAPPα production[65]; Associated with Tau aggregates[66,18]; Associated with Aβ deposits[47] | Cancer[2,15,16]; Synaptic plasticity[105,106]; Learning and Memory[22]; Neuronal death[34,37]; Myelinisation[107]; Neuroinflammation[99]; Atherosclerosis[94,95]; Vascular remodeling[97] |
| **MMP-12** | Macrophages; Microglial cells; Neurons; Astrocytes; Oligodendrocytes; Endothelial cells; Immune cells | Neuroinflammation[92] | Neuroinflammation[108]; Neurodegeneration[109]; Neuroprotection[110,111]; Myelinisation[107]; Inflammation[112,113]; Immunity[114,115]; Atherosclerosis[94,95]; Vascular remodeling[97]; Cancer[116,117] |
| **MMP-13** | Microglial cells; Neurons; Astrocytes; Endothelial cells; Oligodendrocytes; Immune cells | Increased BACE1 levels[49] | Neuroprotection[118]; Cancer[2,15,16]; Vascular remodeling[97]; Atherosclerosis[94,95] |
| **MT1-MMP** | Astrocytes; Neurons; Microglial cells; Endothelial cells; Immune cells | APP metabolism[18,47,77,87,119]; Aβ degradation[69]; Aβ production[77,18]; Associated with Aβ deposits[47] | Neuroinflammation[50]; Glioblastoma[120]; Cancer[121]; Immunity/Inflammation[122,123]; Cell migration[26,124]; Atherosclerosis[94,95]; Vascular remodeling[97] |
| **MT5-MMP** | Neurons; Astrocytes; Immune cells | Aβ production; Neuroinflammation; Synaptic failure; Learning and Memory disabilities[125,126,127]; APP metabolism[18,119,125,126,128]; Associated with Aβ deposits[139] | Synaptogenesis/nervous tissue remodeling and repair[28,129,130]; Inflammation[28,131]; Cancer[132]; Brain tumor[133]; Synaptic activity[134] |

**MT5-MMP, a new potential target in AD**

Among MMPs, MT-MMPs are expressed at a higher level in the CNS, and particularly MT1-MMP, MT3-MMP, MT4-MMP and MT5-MMP.[21] MT5-MMP was identified in 1999 as the most abundant subtype of MT-MMPs in the brain,[133,135] being the only member of the MMP family predominantly expressed at the cerebral level, in embryos and in post-natal pups, in areas of intense neuronal plasticity, suggesting altogether a key role in brain development.[129,136] Its stable post-developmental expression suggests also an implication in neuronal remodeling under physiological and regenerative conditions in adulthood.[28,45,130,137,138]

MT5-MMP has been recently identified has a new player in AD. Indeed, MT5-MMP was first found to be colocalized with amyloid plaques in AD brains, which suggested its participation in the remodeling of injured zones.[139] The generally accepted vision of APP metabolism (see above amyloidogenic/non-amyloidogenic proteolytic pathways) has been recognized as simplistic in recent years in light of the discovery of new APP proteinases.[138,140] As early as 2006, it was demonstrated the capacity of MT1-, MT3- and MT5-MMP to cleave APP in cellulo, at a new site upstream of the β-secretase cleavage site.[119] This work described the formation of APP fragments, different from those classically reported after α- and β-secretase cleavages.

Recently, independent works from two teams have shown that processing of APP by MT5-MMP leads to the generation of new fragments through its η-secretase activity, eventually leading to neurotoxic effects in vitro and in vivo.[125,128]

Willem et al. confirmed in 2015[128] a physiological pathway for APP proteolysis mediated by MT5-MMP at the site previously described by Ahmad et al.,[119] that they called η site $VLAN_{504}$-$M_{505}$ISEPR ($APP_{695}$ numbering). MT5-MMP was capable of displaying η-secretase activity in vivo, unlike its close homologue MT1-MMP.[128] In this new APP proteolytic pathway, MT5-MMP can generate a soluble sAPPη fragment and a residual CTFη fragment anchored to the membrane (Figure 2). CTFη can then be consecutively cleaved either by β-secretase (BACE-1) or by α-secretase (ADAM-10) to release two fragments, a short (92 amino acids) and a slightly longer (108 amino acids) called Aη-β and Aη-α, respectively. In the same study, the authors demonstrated the neurotoxic potential of the Aη-α peptide, which inhibited LTP in primary rat neuronal cultures, while Aη-β had no toxic effect. These data questioned the positive and negative roles generally attributed to α- and β-secretases, respectively. In addition, it was shown that genetic and pharmacological inhibition of BACE-1 (β-secretase) in mice promoted the accumulation of CTFη and Aη-α. As these fragments (at least Aη-α) are potentially neurotoxic, ongoing therapeutic strategies on β-secretase inhibitors should take these data into account as a possible cause of side effects. Although the toxicity of CTFη has not yet been established, they noted that CTFη levels, as those of neurotoxic CTFβ, were enriched in dystrophic neurites in an AD mouse model and in human AD brains.[128]

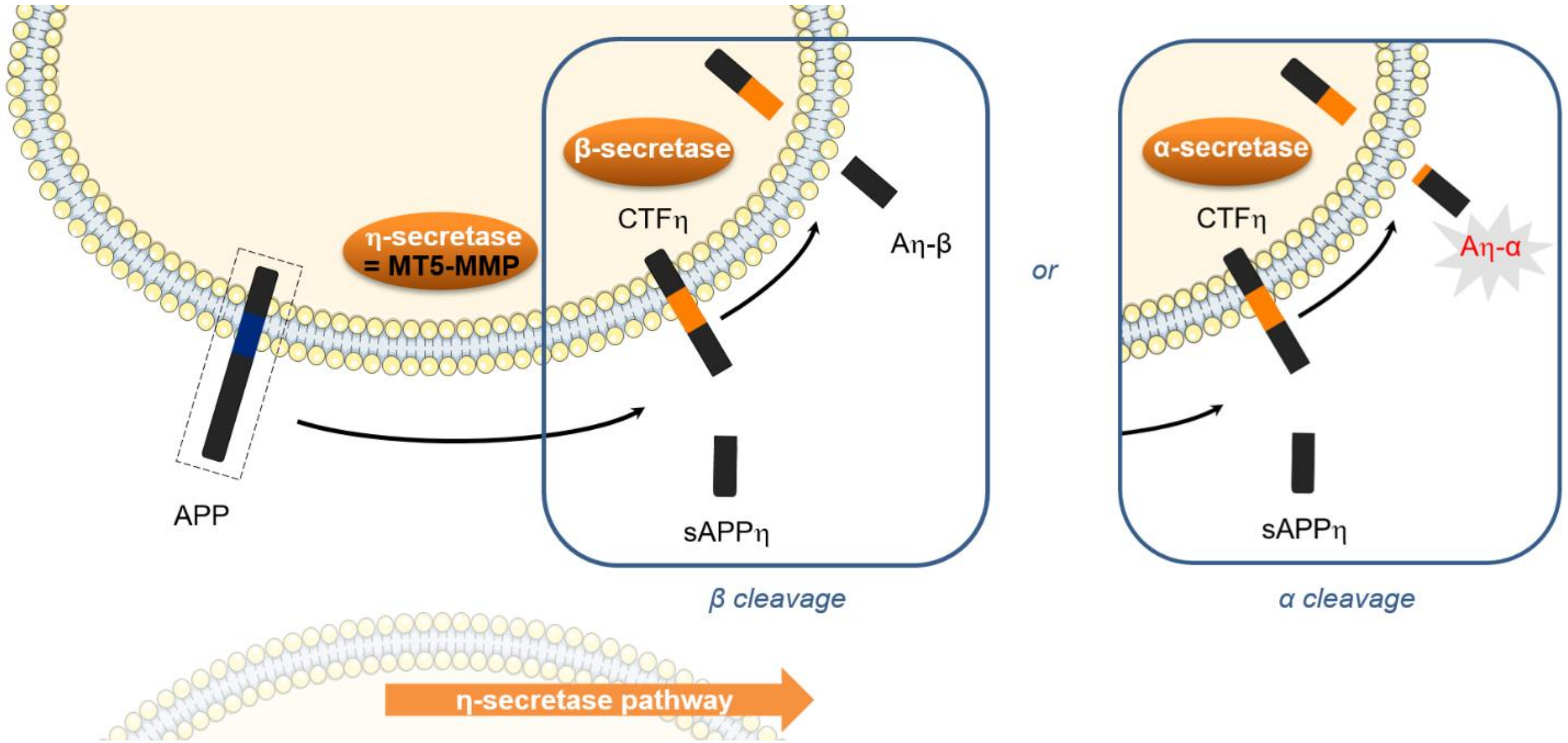

**Figure 2.** Schematic representation of APP processing by η-secretase. Aη-α is neurotoxic, but Aη-β is not.[128]

By the same time, the contribution of MT5-MMP to AD pathogenesis was first unveiled by Baranger et al.[125] They examined the impact of MT5-MMP deficiency in vivo in the 5xFAD mouse model of AD and reported reduced amyloidosis, illustrated by a striking drop in the levels of Aβ (oligomers and amyloid plaques) and CTFβ. Reduced Aβ load was concomitant with durable reduced neuroinflammation and gliosis, as well as the improvement of LTP, and learning and memory. Interestingly, these changes occurred without modifications of the activities of α-, β- and γ- secretases in the cortex and hippocampus. In the same study, they showed in vitro that MT5-MMP can interact and colocalize with APP and also confirmed that MT5-MMP can stimulate Aβ and CTFβ production. Overexpression of MT5-MMP was able to trigger the release of a soluble APP fragment of 95 kDa (sAPP95) in HEK cells which seems to correspond to the sAPPη fragment described by Willem et al..[128] It is noteworthy that sAPP95 levels were significantly reduced in the brain of AD mice deficient for MT5-MMP, therefore further confirming that APP is an in vivo substrate of MT5-MMP.[126]

The pro-amyloidogenic feature of MT5-MMP and its η-secretase activity in the metabolism of APP constitute two pathogenic mechanisms, which may be complementary in AD (Figure 3). It has been suggested, and then confirmed by Baranger et al., that the pro-amyloidogenic function of MT5-MMP could go through a modulation of APP trafficking by facilitating APP sorting in the endosomes, which are a main locus of Aβ formation.[126] In addition, MT5-MMP could also contribute to AD pathogenesis through the activation of pro-inflammatory pathways.[127]

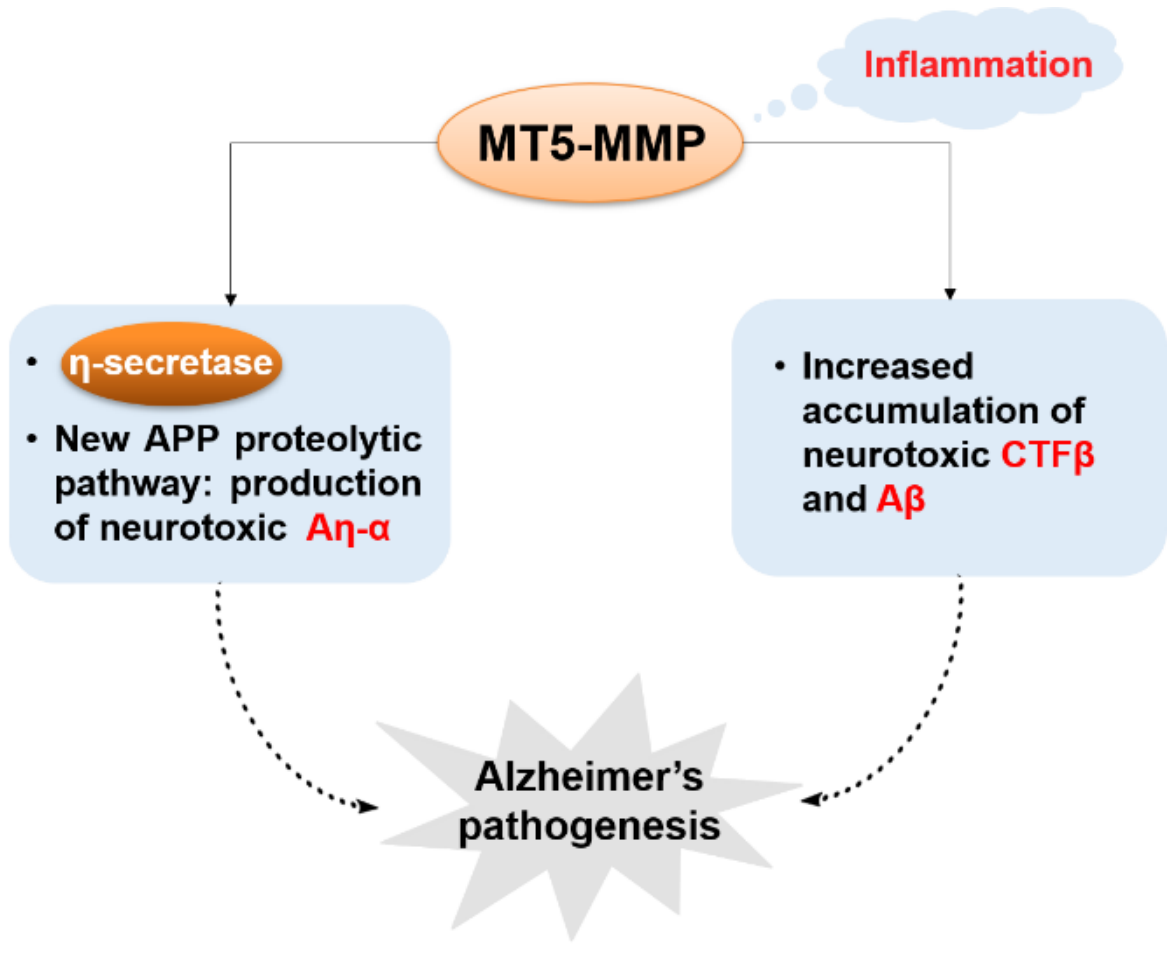

**Figure 3.** Schematic representation of the two newly described disease pathways involving MT5-MMP in the pathogenesis of AD. A third synergistic/complementary pathway would be the role of MT5-MMP in the regulation of inflammatory processes in the CNS, but data are still needed to demonstrate this.

It is also significant to note that MT5-MMP deficient mice are viable and show no detectable overt abnormalities.[131] Keeping in mind that MT5-MMP expression is primarily localized in the nervous system, this information reinforces the idea that MT5-MMP may constitute a relevant therapeutic target in AD, as MT5-MMP inhibitors of therapeutic interest in AD would have limited impact on other organs.

## 5. Major challenges in the design of therapeutic agents targeting MMPs

Given the broad implications of MMPs that have been identified in various neural diseases, including AD, it seems relevant to consider these enzymes as potential new therapeutic targets. However, to develop agents targeting MMPs, it is essential to consider inherent challenges, mainly linked to their broad substrate specificity and high structural homology. In addition, it is important to consider BBB permeability of these agents as CNS targeting drugs. Indeed, despite the high therapeutic potential of MMP inhibitors, almost all clinical trials have failed in the past mainly due to low selectivity and poor target validation.[2,10]

## Broad substrate specificity

As mentioned earlier, MMPs are capable of cleaving a substantial number of extracellular matrix components as well as an increasing number of non-matrix substrates. This confers a wide range of action, and sometimes overlapping substrate specificities to these enzymes, that explains their pleiotropic effects in physiological and pathological conditions.[7,141,142] The search for new MMP substrates using proteomic approaches is still ongoing, with the idea that a thorough knowledge of the substrate repertoire will limit the undesirable side effects associated with unknown substrates. Combining proteomics approaches with studies aimed at characterizing the spatio-temporal expression of MMPs and their substrates should ultimately serve to better understand their biological functions (beneficial and/or detrimental), which is a major challenge in the development of specific MMP inhibitors.[142,143]

## Structural homology

MMPs share a multidomain common structure, with the particularity of catalytic sites that have a high homology in the amino acid sequence.[144] Another characteristic of this proteinase family is the presence of a zinc cation in the catalytic site that is coordinated to three histidine residues within the conserved His-Glu-Xxx-Xxx-His-Xxx-Xxx-Gly-Xxx-Xxx-His motif, and a conserved methionine residue that forms the “Met-turn” region (1,4-β-turn). This conserved zinc-binding environment is what allows MMPs to be included in the Metzincins superfamily.[145] In addition to a zinc atom with catalytic functions, MMPs also share in their active site another zinc atom as well as several calcium ions that act as structural elements in the architecture of the proteinase.[146]

Given the high structural homology between MMPs and the large number of members of this family, a major challenge in the development of therapeutic agents is thus selectivity. In order to introduce selectivity between MMPs, it is necessary to have detailed structural knowledge for each of them. To this end, tertiary MMP structures are of critical importance to design appropriate and selective therapeutic agents. Most of the structures have been solved by X-Ray or nuclear magnetic resonance (NMR) techniques. This mainly concerns catalytic sites, alone or in complex with substrates or various therapeutic agents, although some structures have also been determined for other MMP domains.[3,146,147]

The nomenclature of Schechter and Berger is commonly used to describe catalytic sites of endopeptidases and their substrates.[148] According to this nomenclature, the MMP catalytic site can be divided in several subsites: S1', S2', S3'…S*n*' on the right side of the catalytic zinc cation, i.e., primed side, and S1, S2, S3…S*n* on the left side of the catalytic Zn ion, i.e., unprimed side (Figure 4).

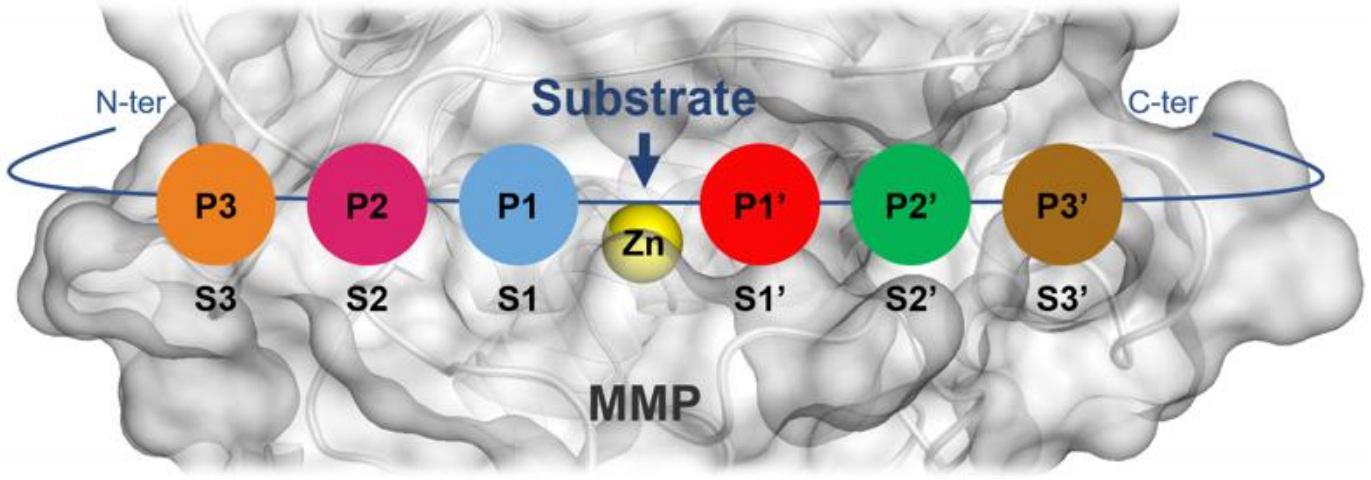

**Figure 4.** Schematic representation of substrate binding in a MMP catalytic site, with the implication of different subsites S*n* of the MMP for the accommodation of various amino acids P*n* of the substrate. In yellow, the zinc cation (Zn), essential for the cleavage of the peptide bond

(indicated by the arrow between P1-P1'). This cartoon has been generated with PyMOL software from the crystal structure of MT3-MMP available in the PDB databank (PDB ID: 1RM8).[149]

By analogy, the corresponding chemical groups on the substrate/analogue that interact with these subsites are named P1, P2…P*n*, and P1', P2'…P*n*'. In MMPs, the non-primed side subsites (mainly S1, S2, S3) have a relatively flat surface and are more exposed to the solvent than the primed-side subsites (mainly S1', S2', S3'), characterized by deep and shallow pockets forming a cleft. The primed-side sites are thus more prone to interactions with the substrates and/or inhibitors and have therefore received more attention in the design of therapeutic agents. Among the subsites in the cleft of the active site, the most interesting is the S1' pocket, which is the closest to the catalytic zinc (Figure 5). This hydrophobic S1′ pocket is the main determinant of enzymatic specificity, as it is critical for substrate recognition. Therefore, it appears to be the most interesting pocket for the design of therapeutic agents because it is the deepest, most flexible and most variable pocket in terms of size, amino acid sequence and shape between MMPs.[150] Nevertheless, other specificity determinants should be considered to improve selectivity, such as the S2' pocket.[144,147,151]

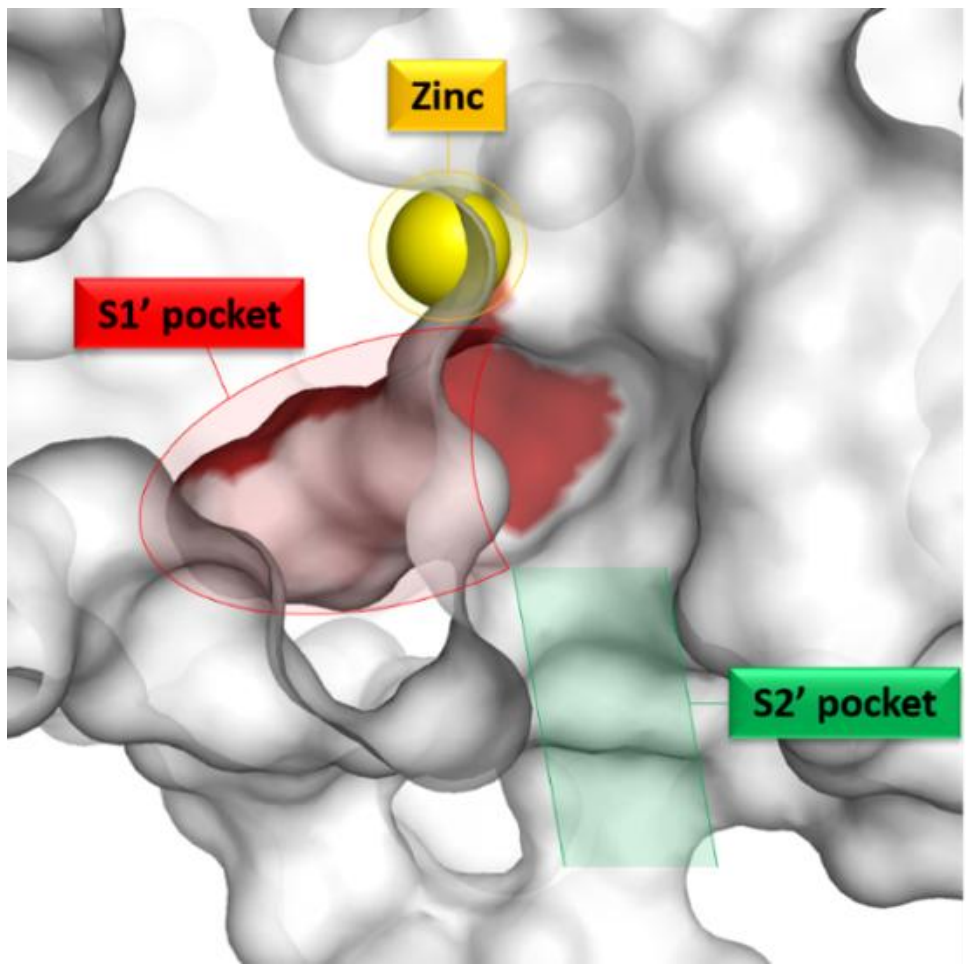

**Figure 5.** Schematic representation of the general catalytic site of MMPs, illustrated by MT3-MMP crystal structure (PDB ID: 1RM8). The S1' pocket (in red), close to Zn cation (in yellow) is deep, hydrophobic and variable in size and shape among the MMPs. The S2' shallow pocket (in green) is very similar among MMPs

and is partially exposed to solvent. This image has been generated with PyMOL software.[149]

A challenge for the coming years will therefore be to characterize the remaining 3D structures of MMPs in order to facilitate structure-based drug design and achieve higher selectivity, either for the catalytic site or for potential exosites.

**BBB permeability**

BBB is a protective element of the brain that provides a defense against pathogenic factors present in the systemic circulation and is therefore crucial for proper synaptic and neuronal functioning. BBB breakdown was demonstrated in several neurodegenerative disorders and particularly in AD, for which the role of its disruption in the pathogenesis of AD is becoming increasingly clear.[72,73,152] In AD, BBB breakdown has been associated to increased BBB permeability, cerebral microbleeds, impaired glucose transport, impaired P-glycoprotein 1 function, CNS leukocyte infiltration, capillary leakages, pericyte and endothelial degeneration, aberrant angiogenesis as well as molecular changes.[72,153] All these alterations lead to the accumulation of toxic molecules in the brain, including Aβ species, that initiate multiple pathways of neurodegeneration. MMPs are versatile enzymes, as seen above, which can have detrimental effects on BBB in various pathological conditions, and thus, could constitute potential therapeutic targets to restore BBB integrity, especially in stroke.[154,155]

The relation between BBB breakdown and brain delivery of neuropharmaceuticals remains a controversial issue. Some say that local BBB disruption could be used by therapeutic agents, and paradoxically by MMP inhibitors themselves, to gain access into CNS, while others postulate that drug delivery requires functionally and structurally healthy BBB.[72,156] CNS delivery of therapeutic agents targeting brain MMPs is indeed another major challenge. Various general strategies for CNS drug delivery can be considered: endogenous cellular mechanisms at the BBB (passive diffusion, carrier-mediated transport, receptor-mediated transcytosis…), intranasal drug delivery, use of nanomedicine, direct injection into CNS or transient opening of the BBB (e.g. focused ultrasound).[72,157,158] More generally, the physicochemical properties of the designed therapeutic agents should be optimized in accordance with the delivery strategy expected in terms of molecular weight, lipophilicity, hydrogen bounding, polar surface area, ionization at physiological pH...[159,160] To conclude, various strategies can be explored to increase brain penetration of potential newly designed MMPs therapeutic agents.

## 6. MMPs inhibition: strategies and evolution

Faced with the increase of functions attributed to MMPs in various pathological conditions, the development of MMP inhibitors has been investigated over time. The general awareness that MMPs assume broader functions than previously thought, has led to the search of the most selective agents as a first step in order to better understand the physiological and pathological role of these enzymes. We will describe the various pharmacological strategies that have been pursued to this end (Figure 6) from the initial idea to block the activity of MMPs by inhibiting their catalytic site. We will thus present the different types of synthetic inhibitors that have been successively designed, using different strategies and with improved results in terms of selectivity. We will finally discuss more recent developments of therapeutic agents targeting exosites, such as monoclonal antibodies (mAb) and protein-based inhibitors. These are designed to target secondary binding sites and to mimic the endogenous regulatory system of MMPs, mainly the TIMPs and the MMP pro-domains, which keep enzymes inactive. As there are just a few examples of therapeutic agents targeting MMPs for brain disorders at the moment, and none for the recently studied MT5-MMP, we will introduce in this section broader examples to present global advancements in the field. We will also include the very few examples that deal with the BBB challenge.

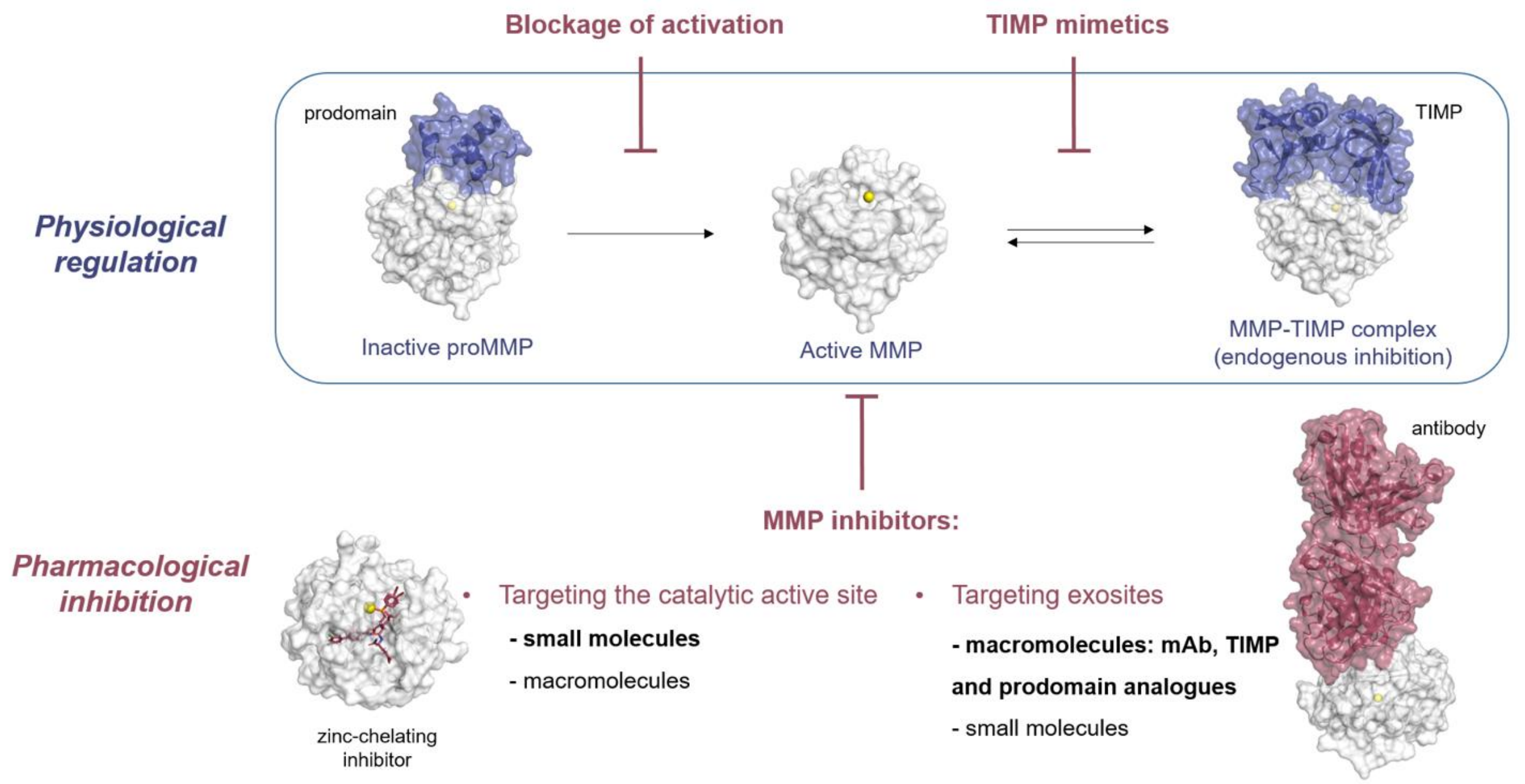

**Figure 6.** Schematic representation of physiological and pharmacological regulation of MMP activity. Physiological regulation is represented in the blue box with proMMP-3 (PDB ID: 1SLM), active MMP-3 (PDB ID: 1CAQ) and the MMP-3/TIMP-1 complex (PDB ID: 1UEA). Pharmacological regulation is presented in red by a zinc-chelating inhibitor **11** complexed with MMP-12 catalytic site (PDB ID: 4GQL) and by GS-5745 exosite-based mAb of MMP-9 (PDB ID: 5TH9). Surface representation of MMP catalytic site with the secondary structure

highlighted in grey, catalytic zinc ion in yellow. The image has been generated with PyMOL software.[149]

**Marketed compounds with anti-MMPs properties**

There are only a few therapeutic drugs on the market that display broad MMP inhibition, and MMPs are generally not even the primary targets. Apart from their known antimicrobial activity, tetracycline-derived compounds, such as Doxycycline **1** and Minocycline, have been shown to inhibit MMPs, particularly collagenases (Figure 7A).[161] Doxycycline hyclate is the only MMP inhibitor with this mechanism that has been approved in periodontal disease by the Food and Drug Administration. It has to be noted that Minocycline is one of the few MMP inhibitors that has been reported to have a good penetration of the BBB.[162] In addition to tetracyclines, bisphosphonates like Alendronate **2** (Figure 7A), which are commonly used for cancer and bone conditions, also appear to inhibit MMPs.[163] Statins, originally designed as cholesterol-lowering drugs (e.g., Simvastatin **3**, Figure 7A), have also more recently shown anti-MMP properties.[164] Finally, Thalidomide **4** (Figure 7A), now available again only through a restricted distribution program, has also shown MMP inhibitory capacity.[165]

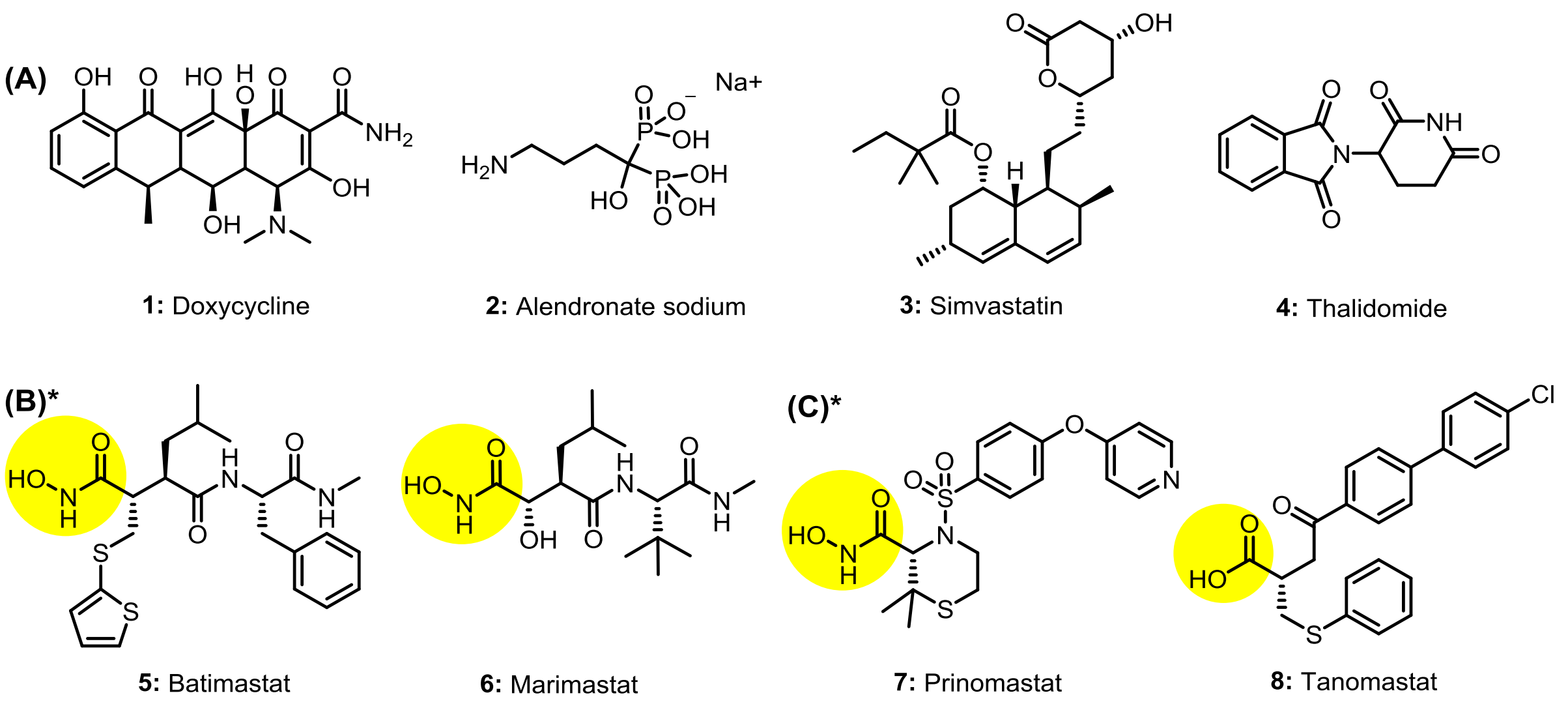

**Figure 7.** Examples of structures of non-selective MMPs inhibitors. (A) Marketed compounds (B) 1st generation: peptides/peptidomimetics (C) 2nd generation: non-peptidomimetics. * Yellow spheres indicate the zinc binding group (ZBG) on designed MMP inhibitors.

### First generation of broad-spectrum peptidic/peptidomimetic inhibitors

In the early design of MMP inhibitors in the 1970s, a substrate-based approach was followed based on the knowledge available at the time. A 1st generation of peptides/peptidomimetics was developed to mimic the structure of their physiological peptide substrates. The requirement for the development of MMP inhibitors was at this time the presence of a zinc binding group (ZBG) on the peptidic scaffold in order to chelate the catalytic zinc atom and block enzyme activity. Hydroxamic acid groups were thus initially chosen for their high potency. Batimastat **5** (also called BB-94, Figure 7B) and many other potent compounds were developed using this approach (for review and complete Structure-Activity Relationships see Whittaker et al.[166]). As represented by Batimastat **5** in Figure 8, these inhibitors were designed to fit the binding cavity of MMPs both through their capacity to chelate zinc, and through their peptidic scaffold that can form hydrogen bonds with the proteinase backbone and distribute its various substituents to the enzyme subsites. However, these peptide-based inhibitors failed in clinical trials for various reasons, including poor oral bioavailability, metabolic instability, lack of MMP selectivity, incomplete knowledge of MMPs biology at the time of the trials, as well as unwanted side effects.[167] Other more orally bioavailable inhibitors, but equally disappointing in clinical trials, were designed by modification of the peptidomimetic backbone. This is the case of Marimastat **6** (also called BB-2516), which possesses a hydroxyl group α-substituent (Figure 7B).[166] However, some of these compounds are still used today as non-selective pharmacological tools in preclinical models.

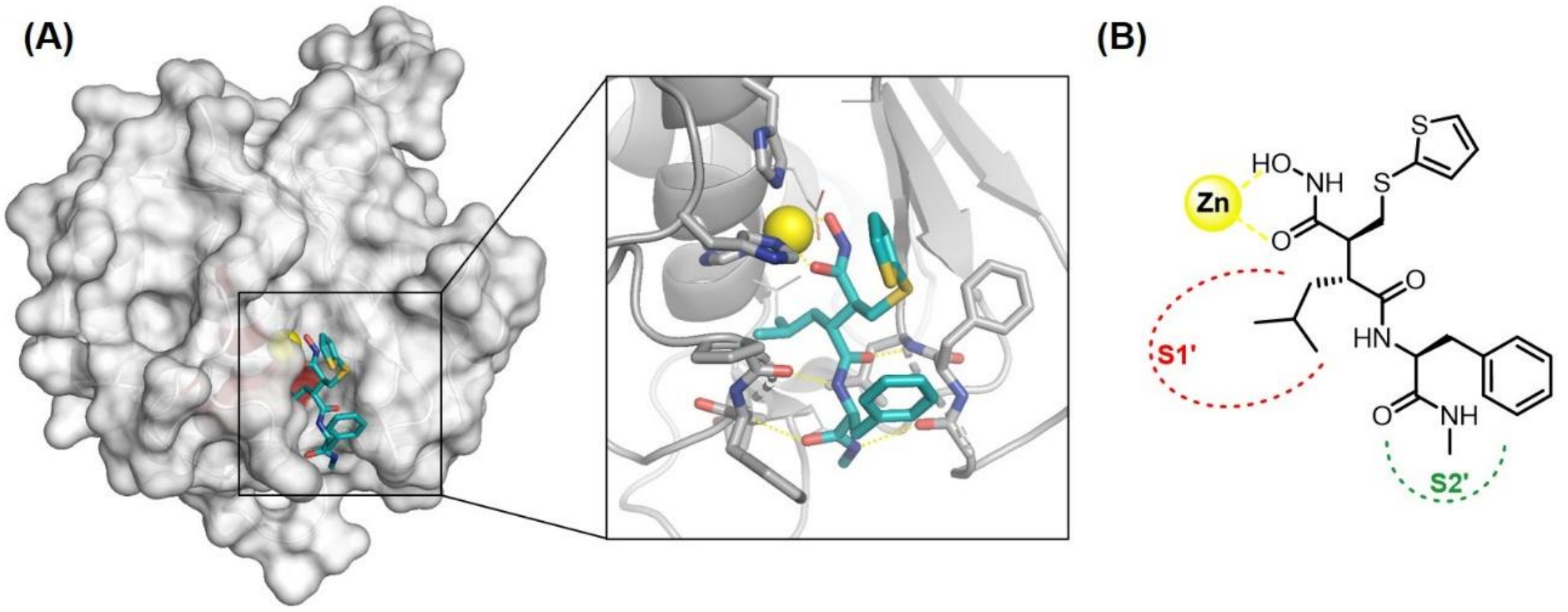

**Figure 8.** (A) Crystal structure of the catalytic domain of MT3-MMP complexed with the non-selective inhibitor Batimastat **5** in blue (PDB ID: 1RM8). Surface representation of the protein with the secondary structure is highlighted in grey, the S1’ pocket in red and the catalytic zinc ion in yellow. The inset shows a zoom of the catalytic site with ligand−protein interactions represented as yellow dashed lines. The image has been generated with PyMOL software.[149] (B) Schematic representation of the binding mode of Batimastat **5** in the catalytic site.

**Second generation of broad-spectrum non-peptidomimetic inhibitors**

Non-peptidyl inhibitors were then explored to target the active site of MMPs and led to a 2nd generation of MMP inhibitors. Different analogues were synthesized, including sulfonamide hydroxamates, like the potent Prinomastat **7** (also called AG3340, Figure 7C).[166] The intense utilization of the hydroxamic acid, plus its lack of specificity, as well as the high competition in the design of MMP inhibitors at that time, led to the search for alternative weaker ZBG such as retrohydroxamates, carboxylates, thiols, phosphorus-based and novel original groups (Figure 9).[151,166,168,169,170]

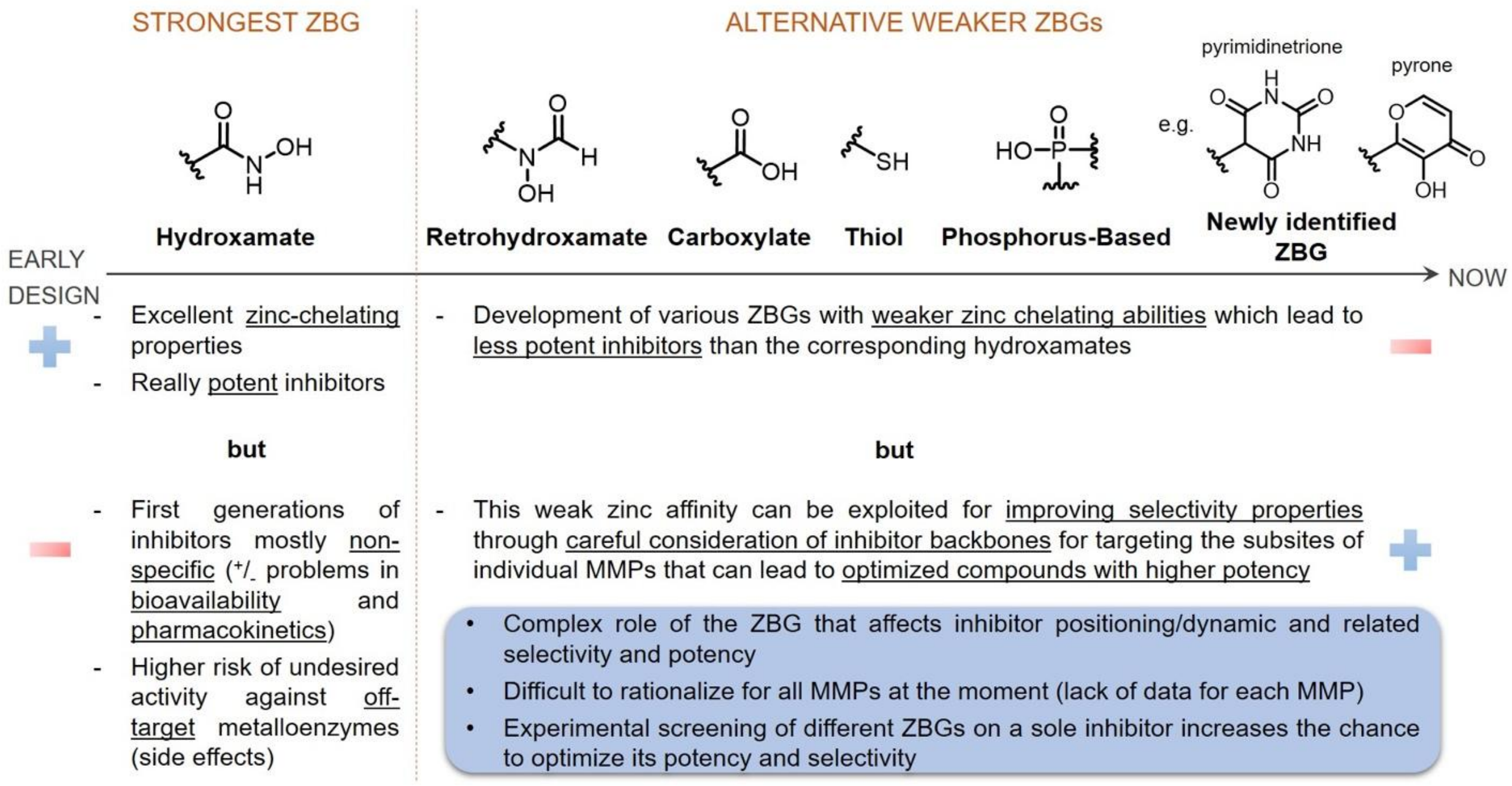

**Figure 9.** Development of various ZBGs over time, represented with their pros and cons.

MMP inhibitors bearing other ZBG were thus developed, like Tanomastat **8** (also called BAY12-9566) possessing a carboxylic acid (Figure 7C).[166] These non-peptidyl inhibitors demonstrated better in vivo pharmacokinetic profiles, but again showed only relatively poor selectivity and therefore failed in clinical trials.[167]

**Third generation of selective inhibitors**

Scientific research over the past two decades has brought new knowledge in the field of MMP inhibitor design and generated technological advances that provide new opportunities, notably in the search of selectivity. The resolution of 3D structures of MMPs is now much more advanced than in the early days of MMPs research, providing key information about MMP structures and MMP-inhibitor interactions that enable the use of structure-based approaches and rational inhibitor design. In addition, this 3$^{rd}$ generation of inhibitors was designed by fully exploiting the presence of various substrate-binding pockets, which surround the catalytically active zinc ion.[151,171] Major selectivity differences were obtained by targeting "selectivity" S1' pockets of MMPs, which have been widely studied and classified as small, medium and large according to the MMP.[150,172] This was achieved in particular through the development of effective technologies and tools based on computer science, high throughput screening (HTS), crystallography, molecular modeling and fragment approach.[173,174] To reach selectivity, scientists are often taking advantage of a combination of these tools. Numerous MMP inhibitors started to emerge with encouraging results in terms of selectivity, in particular for MMP-13 and MMP-12.[168,171,175,176,177]

The sulfonamide moiety has been well studied in the design of MMP inhibitors as an appropriate linker between a substituent oriented to the zinc (mainly a ZBG) and sulfonyl substituents oriented to the S1' pocket. In addition, the sulfonamide linker has shown its ability to do crucial hydrogen bonding in the catalytic site.[178] Successive modifications of this template led to the discovery of various compounds, including N-substituted arylsulfonamide carboxylates like the sugar-based MMP-12 selective inhibitor **9** (Figure 10A).[179,180] Rigidified arylsulfides were also obtained, as illustrated with compound **10**, a selective MMP-12 inhibitor bearing in addition an original N-1-hydroxypiperidine-2,6-dione ZBG (Figure 10A).[181]

In addition, a novel concept for the selective inhibition of MMPs was introduced with the discovery of compound **11** (also called SB-3CT, Figure 10B), a mechanism-based specific inhibitor of MMP-2 and, to a lesser extent, MMP-9 (i.e. suicide substrate).[182] The inhibitor binds deeply in the S1' pocket of the catalytic site with the biphenyl ether moiety while the sulfur atom of the thiirane group coordinates the catalytic zinc. It has been proposed that a nucleophilic attack of a glutamic acid of the active site leads to the opening of the thiirane ring to generate the corresponding thiolate, and forms a covalent ester bond that mimics the proMMP state (slow binding inhibition).[183] Compound **11** and its derivatives have been evaluated in different models of brain diseases thanks to their ability to cross the BBB.[184,185,186,187]

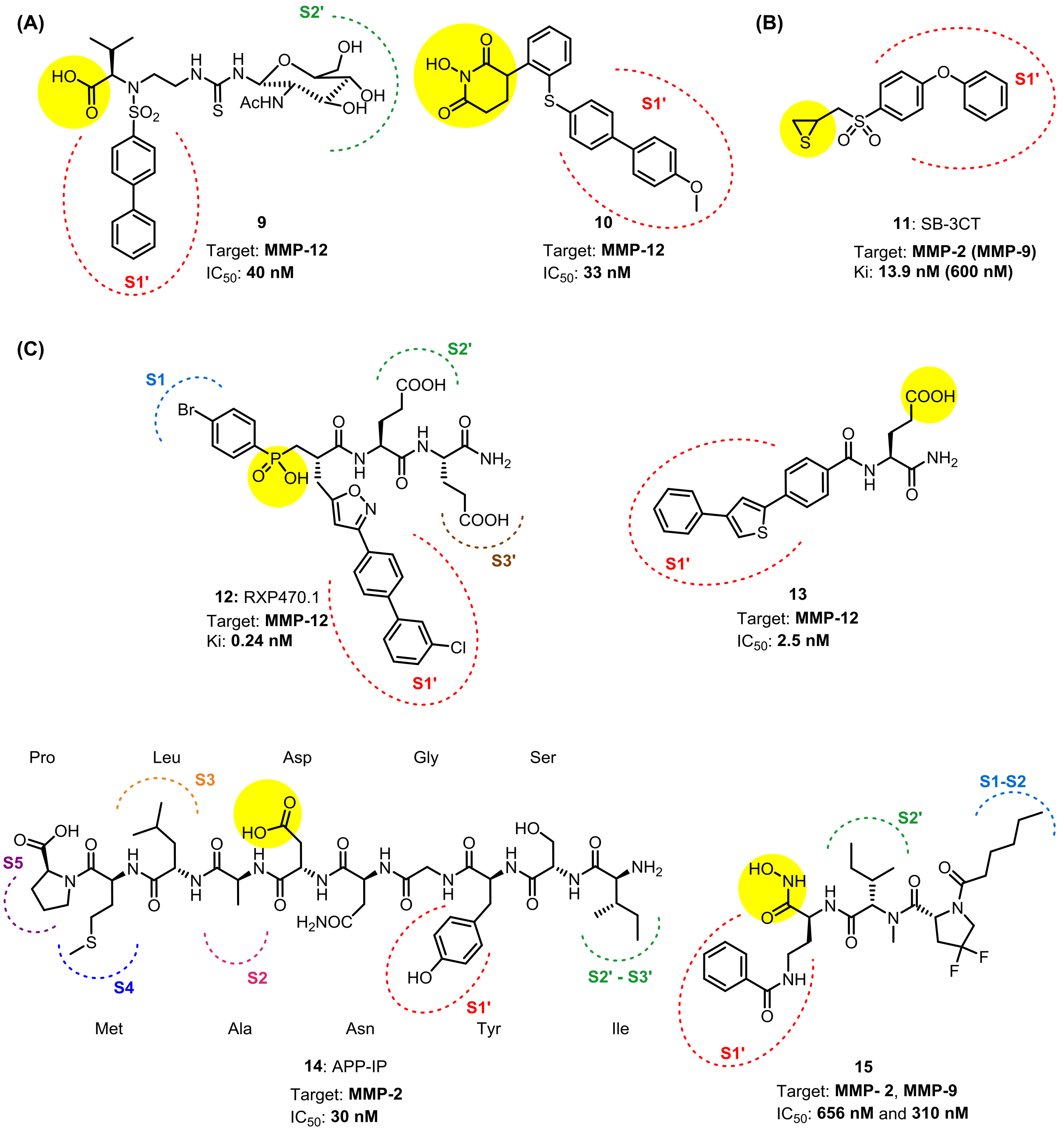

**Figure 10.** Examples of 3rd generation of selective inhibitors with catalytic site interactions represented (Pockets S*n*). Yellow spheres indicate the ZBG. (A) Non-peptidomimetic, (B) mechanism-based and (C) peptidomimetic inhibitors.

Scientists also succeeded more recently to develop selective pseudopeptides, mainly phosphinic pseudopeptides bearing a phosphoryl weak ZBG, which are relatively stable transition-state analogues. Devel et al. designed the highly potent and selective MMP-12 phosphinic inhibitor **12** (also called RXP470.1) by introducing an isoxazole side chain to fill the S1' cavity as well as a Glu-Glu motif to occupy S2' and S3' subsites (Figure 10C).[188,189] An interesting study on this compound revealed the crucial and complex role of the ZBG to modulate either the potency, dynamic or selectivity of the inhibitor.[190] The same group then developed a new generation of non-phosphinic pseudopeptides with modified P1' side chains, like compound **13** (Figure 10C), which were even more selective than **12**.[191,192] In addition to their potential interest as therapeutic agents, some peptidomimetics could be used as probes and pharmacological or diagnostic tools in cellulo or in preclinical models.[193]

Another example of selective inhibitor is the intriguing synthetic APP-IP decapeptide **14**, that interacts with the catalytic site of MMP-2 (Figure 10C). As discussed above, the sequence of APP-IP corresponds to an internal sequence of APP (residues 586–595, $APP_{770}$ numbering) that was identified as the minimal region required for MMP-2 inhibition.[83,84,85] APP-IP **14** specifically inhibits MMP-2 with an $IC_{50}$ value of 30 nM while sparing MMPs like MT1-MMP, MMP-3, -7 and -9, with $IC_{50}$ values between $10^3$ and $10^4$ higher.[83,84,85]

The hydroxamate-based gelatinases inhibitor BBB-permeable **15** has come to light in a paper just published by Bertran et al. (Figure 10C).[194] They have successfully developed through a multidisciplinary approach a selective potent inhibitor of MMP-2 and MMP-9 with potential for the treatment of CNS, as suggested by its ability to cross in vitro and in vivo the BBB. They have applied a structure-based drug design approach, after an initial in silico screening, to optimize the potency, the selectivity as well as the permeability of the compounds. When considering multiple parameters since the beginning, it is so feasible to overcome several main challenges in the development of MMP inhibitors, namely selectivity and BBB-permeability.

**Fourth generation of exosite-based inhibitors**

In recent years, inhibitors that target exosites have also been studied to help unravel the complex issue of the selectivity.[195] Some of the least conserved regions among MMPs, known as hot spots, have been specifically identified for each MMP, opening up new opportunities to develop an even more selective 4$^{th}$ generation of exosite-based inhibitors.[196] Secondary binding sites have been found, notably for MMP-13 in the catalytic domain, for MT1-MMP in the hemopexin-like domain as well as in the catalytic site, and for MMP-2 and MMP-9 in the collagen binding domain.[197] Very potent peptides and small molecules have thus been developed, in particular for MMP-13.[198]

To circumvent broad-spectrum MMP inhibition or the inhibition of other metalloenzymes, the development of compounds that do not interact with the zinc, without any ZBG, was considered in the meantime.[168] In this context, an interesting approach allowed the identification of ZBG-free highly selective and potent MMP-13 inhibitors. As a result of initial HTS experiments, a first lead capable of binding the S1' specificity subsite, also called S1" pocket, has been identified.[199] The latter, as mentioned earlier is a well-known exosite of the MMP-13 catalytic site. Studies based on the relationship between structure and activity, as well as the crystallization of a complex formed by one of the inhibitors and MMP-13, allowed to use comparative structural analyses to design a highly potent zinc-chelating inhibitor, compound **16**, which was non-selective (Figure 11A).[200,201,202] Another structure-guided molecular design led in a second phase to the synthesis of inhibitor **17**, which was in this case selective for MMP-13 through the removal of the ZBG (Figure 11A).[202]

In another example, Nara et al. also first identified a hit compound by HTS, which was then co-crystallized with the catalytic site of MMP-13. Next, they applied a structure-based drug design approach to target the deep S1′ pocket and the unique MMP-13 adjacent side pocket S1’’. This led to the identification of the highly potent and selective MMP-13 inhibitor **18** lacking the ZBG, but which was limited by its pharmacokinetic profile (Figure 11A).[203] Nara et al. optimized compound **18** by introducing a 1,2,4-triazol-3-yl group as ZBG, resulting in MMP-13 inhibitor **19** with excellent potency and selectivity, as well as improved favorable pharmacokinetic properties (Figure 11A).[204] Despite some concerns about off-targets and as depicted with this example, selective MMP inhibitors bearing a ZBG can be successfully developed by carefully examining the inhibitor scaffold to target specific subsites of individual MMPs.[168]

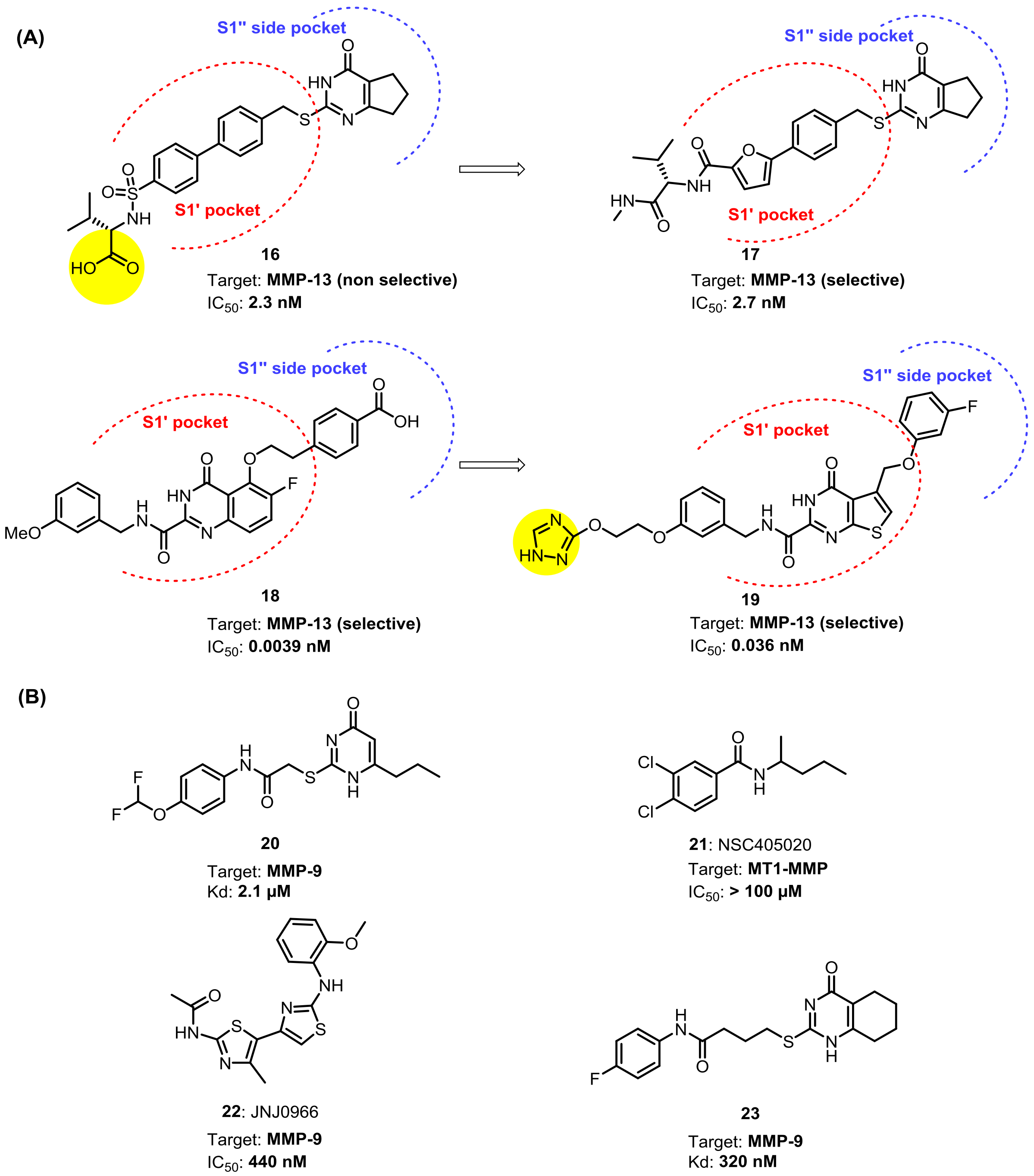

**Figure 11.** Examples of the 4th generation exosite-based MMP inhibitors. (A) MMP-13 inhibitors with protein-inhibitor interactions represented (MMP pockets S*n*' are visualized and yellow spheres indicate the ZBG). (B) Other exosite-based MMP inhibitors.

Inhibitors targeting exosites are beginning to emerge also for other MMPs. The in silico docking approach, followed by an experimental assay, identified a selective inhibitor of the hemopexin-like domain of MMP-9 (exosite), compound **20**, which reduced cancer cell migration and proliferation without modulating MMP-9 catalytic activity (Figure 11B).[205] The hemopexin domain of MT1-MMP was also exploited for exosite-based inhibitor development. In this case, a virtual ligand screening led to the identification of compound NSC405020 **21**, a small molecule that binds the MT1-MMP hemopexin domain, showed antitumor efficacy in vitro and impaired MT1-MMP homodimerization, but not proteolytic activity (Figure 11B).[206] An unprecedented pharmacological approach was developed with compound JNJ0966 **22**, which selectively inhibits the conversion of inactive proMMP-9 into its active form by MMP-3, but does not inhibit active MMP-9 nor MMP-1, MMP-3 and MT1-MMP (Figure 11B). Compound **22** was shown to interact directly with an exosite located near the proMMP-9 cleavage site inside the prodomain and to reduce experimental autoimmune encephalomyelitis in vivo, thus validating this pharmacological approach.[207] Another small molecule inhibitor **23** that targets specifically the hemopexin domain of proMMP-9 (not proMMP-2, nor proMT1-MMP) has been discovered and shown to block in vivo cancer cell invasion and angiogenesis. This compound was designed by in silico studies, based on compound **20**, and demonstrated its ability to prevent proMMP-9 homodimerization which is critical for cell migration.[208]

The search for small organic molecules targeting the active site remains the most followed strategy at present. However, other strategies to control MMP activities via alternative exosites offer significant opportunities in MMP drug design.

**Development of macromolecules**

Given the challenges of developing selective MMP inhibitors that target the catalytic site, exosites outside the catalytic site are now also being explored, as referred above.[195,196,197] However, the relatively small size of peptides and/or small molecules limits the possibilities of targeting more specific interactions in more distant exosites. This goal can be more easily reached with emerging protein-based agents.[209,210,211] Indeed, highly selective engineered mAb (or nanobodies) directed against different MMPs, including MMP-2, -8, -9, -13 and MT1-MMP, were developed targeting either exosites or the catalytic site.[14,212,213,214] Another approach is to try to mimic the endogenous regulatory system of MMPs with endogenous-like inhibitors, i.e., TIMPs or pro-domain analogues, which are non-toxic and well-known.[209] In this context, protein engineering tools have been used to design a family of N-terminally modified TIMPs.[14] For example, a fusion protein consisting of the synthetic decapeptide APP-IP **14** and TIMP-2 was designed and demonstrated $10^6$ greater inhibitory activity against MMP-2 ($IC_{50}$ 0.7 pM), compared to MMP-1, -3, -7, -8, -9, or MT1-MMP ($IC_{50}$ > 1μM).[86] This protein-based inhibitor has shown its ability to interact both with the catalytic site and the hemopexin-like domain of MMP-2.[86]

Although outstanding selectivity can be obtained with this type of macromolecules, a number of limitations have to be taken into account with these therapeutic agents, mainly high cost of production, difficulties in distribution/stability/clearance, permeability problems across the biological membranes, as well as immunogenicity.[209]

## 7. Conclusions and Perspectives

We have presented in this article the complex roles of MMPs in physiological and pathological processes, as well as their potential therapeutic interest. Various types of therapeutic agents have been developed over time to address the challenge of making increasingly selective MMP modulators in order to decipher MMP specific functions and to limit side effects. Small molecules, peptides or protein-based inhibitors are part of the repertoire of new molecules, and the discovery of specific exosites broadens the range of possibilities to generate new highly selective compounds. Some of the available agents demonstrate already their efficacy in in vivo models of several diseases such as atherosclerosis with MMP-12,[215] osteoarthritis with MMP-13,[203,216,217,218] cancer[80,206,219,220,221] and rheumatoid arthritis[222] with MT1-MMP. Some of them are currently involved in clinical trials, including Andecaliximab (GS-5745), a mAb targeting MMP-9 for glioblastoma (NCT03631836) or FP-025 (Forsee Pharmaceutical), a MMP-12 non-hydroxamate inhibitor for allergic asthma (NCT03858686).

However, even if selectivity remains the primary objective at the moment, it is important to note the possible synergy of targeting more than one MMP for better efficacy through a polypharmacological approach. Along this line, we could cite ZHAWOC6941, a non-hydroxamate dual inhibitor of MMP-7 and MMP-13, two validated targets in cancer,[223] or ZHAWOC7726, a TIMP peptidomimetic, with a good selectivity towards the anti-cancer targets MMP-13, MMP-2 and MMP-9.[224] The use of multitarget agents with pleiotropic actions might highlight future trends in the field of neurodegenerative diseases, and especially AD, a complex multicausal disorder. This would imply targeting more than one MMP (among MMP-1, -2, -3, -9, -12, -13, MT1-MMP or MT5-MMP) with pan-specific inhibitors,[225] or combining targeting of one MMP with another target of interest in AD, e.g. cholinesterases.[226] Beyond specificity considerations, targeting of MMPs in AD must take into account their generally wide distribution, with the exception of MT5-MMP that is primarily expressed in the nervous system. Thus, inhibition of MT5-MMP may offer some advantage in terms of brain targeting relative to other MMP homologues. Based on progress in structural biology, design of druggable inhibitors towards MT5-MMP appears as a realistic achievement and makes it altogether a peculiarly promising target among MMPs for a positive impact in AD. In any event, therapeutic strategies should ideally consider delivery of active compounds with molecular conjugates that optimize access to the brain via the BBB. The prospects of future anti-MMP therapies in AD should also integrate the possibility that fragments resulting from substrate cleavage may be targets *per se* and their actions more easily modulated than the proteolytic processes that generate them. Increasing knowledge on the nature of MMP substrates, many of them inflammatory mediators, should open new avenues for therapeutic intervention in this domain as well. Finally, the chronic nature of AD and the fact that the pathological processes precede symptoms by 10-15 years makes it necessary to consider when to target a given MMP based on its pattern of expression over time during disease progression. A better knowledge of the spatio-temporal distribution of MMPs in the brain should therefore increase the chances of accurately defining the time frame of intervention. Overall, a thorough knowledge of the pathophysiological processes underlying AD appears to be an inevitable path to designing more effective therapeutic strategies, including those based on MMP inhibition.

## AUTHOR INFORMATION

### Corresponding Author

*Patrick Dallemagne: phone, +33-231566813; e-mail, patrick.dallemagne@unicaen.fr.

**ACKNOWLEDGMENT**

The PhD project of PZ (MT5-MMP « Design, synthesis and biological evaluation of novel MT5-MMP inhibitors with potential interest for Alzheimer's disease ») is co-financed by the European Union under the 'FEDER-FSE 2014-2020' operational program and supported by France Alzheimer. This work is co-funded by Région Normandie.

This work was supported by funding from the CNRS and Aix-Marseille Université, by France Alzheimer grants to SR and PD, and by the French National Research Agency (ANR-MAD5) to SR, as part of the second "Investissements d'Avenir" program. KB was granted by the Excellence Initiative of Aix-Marseille Université - A*MIDEX, a French "Investissements d'Avenir" program.

Some pictures were created in part from SMART (Servier Medical Art) images, licensed under a Creative Common Attribution 3.0 Generic License. http://smart.servier.com/

**ABBREVIATIONS**

Aβ, amyloid-β peptide; ADAM, a disintegrin and metalloprotease domain; AICD, APP Intracellular Domain; APP, amyloid precursor protein; CTF, C-terminal fragment; CNS, central nervous system; mAb, monoclonal antibody; MMP, matrix metalloproteinase; MT-MMP, membrane-type MMP; sAPP, soluble APP; TIMP, tissue inhibitor of matrix metalloproteinase; ZBG, zinc binding group

**BIOGRAPHIES**:

**Pauline Zipfel** obtained her PharmD degree in 2017 from the University of Strasbourg (France). The same year, she joined the group of Prof. P. Dallemagne at the "Centre d'Etudes et de Recherche sur le Médicament de Normandie" (CERMN) laboratory at the University of Caen Normandy (France) for a Medicinal Chemistry PhD position. She is currently working in her final year to complete her PhD and her research is investigating the design, synthesis and biological evaluation of first-in-class small molecule inhibitors of MT5-MMP, as an innovative strategy for the treatment of Alzheimer's disease. She also decided to join in 2019 the European Federation for Medicinal Chemistry (EFMC) communication team to promote cooperation between medicinal chemists in Europe and around the World.

**Christophe Rochais** received his Engineer Diploma in Chemistry from ENSCMu in 2002, and his PhD in the group or Prof. S. Rault from the University of Caen Basse-Normandie (2002-2005). After a post-doctoral fellowship in the University of Nottingham, he was appointed Lecturer in Organic Chemistry in the School of Pharmacy at the University of Caen Normandie in 2007 and since 2014 he assumed the position of Professor. His research interests include medicinal chemistry program in the field of enzymatic inhibition and GPCR modulation to develop pharmacological tools and bioactive compounds. He is leading a research group dedicated to the development of pleiotropic agents of interest for Alzheimer's disease and has been recently appointed as a member of the French National Academy of Pharmacy.

**Kévin Baranger** received his Ph.D. in protein biochemistry within the group of Professor Thierry Moreau from the University of Tours in 2008. After completing a postdoctoral fellowship at the NICN lab of Aix-Marseille University (2009-2017), under the supervision of Dr Santiago Rivera, he was appointed CNRS researcher at the Institute of NeuroPathophysiology UMR7051 at Aix-Marseille University. He is a specialist of proteinases/inhibitors in pathophysiologic processes. Since 2009, his work has been dedicated to understanding the role of MT5- and MT1-MMP in Alzheimer's disease pathogenesis and to the development of new therapeutic strategies.

**Santiago Rivera** received his PhD in Neuroscience from the University of Barcelona in 1990. After a post-doctorate at the University of California (Irvine) and Inserm (Paris), he was appointed researcher at the CNRS in 1998 and became head of the Neural Plasticity and Degeneration group in 2002. In 2010, he was appointed Research Director at the CNRS, and since 2018 he is Deputy Director of the Institute of Neuropathophysiology at the University of Aix-Marseille. He is a member of the scientific advisory board of various Alzheimer's disease foundations. Dr. Rivera's work has focused primarily on the study of the pathophysiological mechanisms underlying neurodegenerative disorders, in particular the role of MMPs and their natural inhibitors, and on the search for innovative therapeutic approaches in Alzheimer's disease.

**Patrick Dallemagne** studied pharmacy at the University of Caen and obtained his PharmD in 1983. He received his PhD in Medicinal Chemistry in 1988 and his Habilitation Diploma in 1990. He became Associate Professor of Medicinal Chemistry at the University of Caen Normandie in 1991 and was received Professor in 1999. He is currently head of the Centre d'Etudes et de Recherche sur le Médicament de Normandie, where he developed a lot of novel compounds with therapeutic interest in oncology and neurosciences areas. Actually, his group works on programs concerning novel Multi-Target Directed Ligands and recently succeeded in the design of donecopride, a first dual 5-HT$_4$R agonist/AChEI currently in preclinical trials against Alzheimer's disease. He is member of the French National Academy of Pharmacy.

REFERENCES

(1) Patterson, C. World Alzheimer Report 2018. The State of the Art of Dementia Research: New Frontiers. *London Alzheimer's Dis. Int.* **2018**. https://www.alz.co.uk/research/WorldAlzheimerReport2018.pdf (accessed Feb 8, 2020

(2) Vandenbroucke, R. E.; Libert, C. Is There New Hope for Therapeutic Matrix Metalloproteinase Inhibition? *Nat. Rev. Drug Discov.* **2014**, *13* (12), 904–927. https://doi.org/10.1038/nrd4390.

(3) Maskos, K. Crystal Structures of MMPs in Complex with Physiological and Pharmacological Inhibitors. *Biochimie* **2005**, *87* (3–4), 249–263. https://doi.org/10.1016/j.biochi.2004.11.019.

(4) Nagase, H.; Visse, R.; Murphy, G. Structure and Function of Matrix Metalloproteinases and TIMPs. *Cardiovasc. Res.* **2006**, *69* (3), 562–573. https://doi.org/10.1016/j.cardiores.2005.12.002.

(5) Rivera, S.; Khrestchatisky, M.; Kaczmarek, L.; Rosenberg, G. A.; Jaworski, D. M.

Metzincin Proteases and Their Inhibitors: Foes or Friends in Nervous System Physiology? *J. Neurosci.* **2010**, *30* (46), 15337–15357. https://doi.org/10.1523/JNEUROSCI.3467-10.2010.

(6) Kumar, G. B.; Nair, B. G.; Perry, J. J. P.; Martin, D. B. C. Recent Insights into Natural Product Inhibitors of Matrix Metalloproteinases. *Medchemcomm* **2019**, *10* (12), 2024–2037. https://doi.org/10.1039/C9MD00165D.

(7) Sternlicht, M. D.; Werb, Z. How Matrix Metalloproteinases Regulate Cell Behavior. *Annu. Rev. Cell Dev. Biol.* **2001**, *17* (1), 463–516. https://doi.org/10.1146/annurev.cellbio.17.1.463.

(8) Page-McCaw, A.; Ewald, A. J.; Werb, Z. Matrix Metalloproteinases and the Regulation of Tissue Remodelling. *Nat. Rev. Mol. Cell Biol.* **2007**, *8* (3), 221–233. https://doi.org/10.1038/nrm2125.

(9) Gaffney, J.; Solomonov, I.; Zehorai, E.; Sagi, I. Multilevel Regulation of Matrix Metalloproteinases in Tissue Homeostasis Indicates Their Molecular Specificity in Vivo. *Matrix Biol.* **2015**, *44–46* (Ldl), 191–199. https://doi.org/10.1016/j.matbio.2015.01.012.

(10) Baranger, K.; Rivera, S.; Liechti, F. D.; Grandgirard, D.; Bigas, J.; Seco, J.; Tarrago, T.; Leib, S. L.; Khrestchatisky, M. *Endogenous and Synthetic MMP Inhibitors in CNS Physiopathology*; Progress in Brain Research, 2014; Chapter 214, pp 313–351. https://doi.org/10.1016/B978-0-444-63486-3.00014-1.

(11) Brew, K.; Dinakarpandian, D.; Nagase, H. Tissue Inhibitors of Metalloproteinases: Evolution, Structure and Function. *Biochim. Biophys. Acta - Protein Struct. Mol. Enzymol.*

**2000**, *1477* (1–2), 267–283. https://doi.org/10.1016/S0167-4838(99)00279-4.

(12) Stamenkovic, I. Extracellular Matrix Remodelling: The Role of Matrix Metalloproteinases. *J. Pathol.* **2003**, *200* (4), 448–464. https://doi.org/10.1002/path.1400.

(13) Parks, W. C.; Wilson, C. L.; López-Boado, Y. S. Matrix Metalloproteinases as Modulators of Inflammation and Innate Immunity. *Nat. Rev. Immunol.* **2004**, *4* (8), 617–629. https://doi.org/10.1038/nri1418.

(14) Levin, M.; Udi, Y.; Solomonov, I.; Sagi, I. Next Generation Matrix Metalloproteinase Inhibitors — Novel Strategies Bring New Prospects. *Biochim. Biophys. Acta - Mol. Cell Res.* **2017**, *1864* (11), 1927–1939. https://doi.org/10.1016/j.bbamcr.2017.06.009.

(15) Overall, C. M.; Kleifeld, O. Validating Matrix Metalloproteinases as Drug Targets and Anti-Targets for Cancer Therapy. *Nat. Rev. Cancer* **2006**, *6* (3), 227–239. https://doi.org/10.1038/nrc1821.

(16) Fingleton, B. Matrix Metalloproteinases as Valid Clinical Targets. *Curr. Pharm. Des.* **2007**, *13* (3), 333–346. https://doi.org/10.2174/138161207779313551.

(17) Brkic, M.; Balusu, S.; Libert, C.; Vandenbroucke, R. E. Friends or Foes: Matrix Metalloproteinases and Their Multifaceted Roles in Neurodegenerative Diseases. *Mediators Inflamm.* **2015**, *2015*, 1–27. https://doi.org/10.1155/2015/620581.

(18) Rivera, S.; García-González, L.; Khrestchatisky, M.; Baranger, K. Metalloproteinases and Their Tissue Inhibitors in Alzheimer's Disease and Other Neurodegenerative Disorders. *Cell. Mol. Life Sci.* **2019**, *76* (16), 3167–3191. https://doi.org/10.1007/s00018-019-03178-

2.

(19) Dzwonek, J.; Rylski, M.; Kaczmarek, L. Matrix Metalloproteinases and Their Endogenous Inhibitors in Neuronal Physiology of the Adult Brain. *FEBS Lett.* **2004**, *567* (1), 129–135. https://doi.org/10.1016/j.febslet.2004.03.070.

(20) Agrawal, S.; Lau, L.; Yong, V. MMPs in the Central Nervous System: Where the Good Guys Go Bad. *Semin. Cell Dev. Biol.* **2008**, *19* (1), 42–51. https://doi.org/10.1016/j.semcdb.2007.06.003.

(21) Brzdak, P.; Nowak, D.; Wiera, G.; Mozrzymas, J. W. Multifaceted Roles of Metzincins in CNS Physiology and Pathology: From Synaptic Plasticity and Cognition to Neurodegenerative Disorders. *Front. Cell. Neurosci.* **2017**, *11*, 1–22. https://doi.org/10.3389/fncel.2017.00178.

(22) Beroun, A.; Mitra, S.; Michaluk, P.; Pijet, B.; Stefaniuk, M.; Kaczmarek, L. MMPs in Learning and Memory and Neuropsychiatric Disorders. *Cell. Mol. Life Sci.* **2019**, *76* (16), 3207–3228. https://doi.org/10.1007/s00018-019-03180-8.

(23) Sanz, R. L.; Ferraro, G. B.; Kacervosky, J.; Salesse, C.; Gowing, E.; Hua, L.; Rambaldi, I.; Beaubien, F.; Holmbeck, K.; Cloutier, J. F.; Levesque M.; Murai K.; Fournier A. E. MT3-MMP Promotes Excitatory Synapse Formation by Promoting Nogo-66 Receptor Ectodomain Shedding. *J. Neurosci.* **2018**, *38* (3), 518–529. https://doi.org/10.1523/JNEUROSCI.0962-17.2017.

(24) Powell, M. A.; Black, R. T.; Smith, T. L.; Reeves, T. M.; Phillips, L. L. Matrix Metalloproteinase 9 and Osteopontin Interact to Support Synaptogenesis in the Olfactory

Bulb after Mild Traumatic Brain Injury. *J. Neurotrauma* **2019**, *36* (10), 1615–1631. https://doi.org/10.1089/neu.2018.5994.

(25) Ogier, C.; Bernard, A.; Chollet, A.-M.; LE Diguardher, T.; Hanessian, S.; Charton, G.; Khrestchatisky, M.; Rivera, S. Matrix Metalloproteinase-2 (MMP-2) Regulates Astrocyte Motility in Connection with the Actin Cytoskeleton and Integrins. *Glia* **2006**, *54* (4), 272–284. https://doi.org/10.1002/glia.20349.

(26) Ould-Yahoui, A.; Sbai, O.; Baranger, K.; Bernard, A.; Gueye, Y.; Charrat, E.; Clément, B.; Gigmes, D.; Dive, V.; Girard, S. D.; Féron, F.; Khrestchatisky, M.; Rivera, S. Role of Matrix Metalloproteinases in Migration and Neurotrophic Properties of Nasal Olfactory Stem and Ensheathing Cells. *Cell Transplant.* **2013**, *22* (6), 993–1010. https://doi.org/10.3727/096368912X657468.

(27) Gueye, Y.; Ferhat, L.; Sbai, O.; Bianco, J.; Ould-Yahoui, A.; Bernard, A.; Charrat, E.; Chauvin, J.-P.; Risso, J.-J.; Féron, F.; Rivera, S.; Khrestchatisky, M. Trafficking and Secretion of Matrix Metalloproteinase-2 in Olfactory Ensheathing Glial Cells: A Role in Cell Migration? *Glia* **2011**, *59* (5), 750–770. https://doi.org/10.1002/glia.21146.

(28) Warren, K. M.; Reeves, T. M.; Phillips, L. L. MT5-MMP, ADAM-10, and N-Cadherin Act in Concert To Facilitate Synapse Reorganization after Traumatic Brain Injury. *J. Neurotrauma* **2012**, *29* (10), 1922–1940. https://doi.org/10.1089/neu.2012.2383.

(29) Huntley, G. W. Synaptic Circuit Remodelling by Matrix Metalloproteinases in Health and Disease. *Nat. Rev. Neurosci.* **2012**, *13* (11), 743–757. https://doi.org/10.1038/nrn3320.

(30) Larsen, P. H. Myelin Formation during Development of the CNS Is Delayed in Matrix

Metalloproteinase-9 and -12 Null Mice. *J. Neurosci.* **2006**, *26* (8), 2207–2214. https://doi.org/10.1523/JNEUROSCI.1880-05.2006.

(31) Hsu, J.-Y. C.; McKeon, R.; Goussev, S.; Werb, Z.; Lee, J.-U.; Trivedi, A.; Noble-Haeusslein, L. J. Matrix Metalloproteinase-2 Facilitates Wound Healing Events That Promote Functional Recovery after Spinal Cord Injury. *J. Neurosci.* **2006**, *26* (39), 9841–9850. https://doi.org/10.1523/JNEUROSCI.1993-06.2006.

(32) Larsen, P. H.; Wells, J. E.; Stallcup, W. B.; Opdenakker, G.; Yong, V. W. Matrix Metalloproteinase-9 Facilitates Remyelination in Part by Processing the Inhibitory NG2 Proteoglycan. *J. Neurosci.* **2003**, *23* (35), 11127–11135. https://doi.org/10.1523/JNEUROSCI.23-35-11127.2003.

(33) Zhang, J. W.; Deb, S.; Gottschall, P. E. Regional and Age-Related Expression of Gelatinases in the Brains of Young and Old Rats after Treatment with Kainic Acid. *Neurosci. Lett.* **2000**, *295* (1–2), 9–12. https://doi.org/10.1016/S0304-3940(00)01582-2.

(34) Jourquin, J.; Tremblay, E.; Decanis, N.; Charton, G.; Hanessian, S.; Chollet, A.-M.; Le Diguardher, T.; Khrestchatisky, M.; Rivera, S. Neuronal Activity-Dependent Increase of Net Matrix Metalloproteinase Activity Is Associated with MMP-9 Neurotoxicity after Kainate. *Eur. J. Neurosci.* **2003**, *18* (6), 1507–1517. https://doi.org/10.1046/j.1460-9568.2003.02876.x.

(35) Gasche, Y. Matrix Metalloproteinases and Diseases of the Central Nervous System with a Special Emphasis on Ischemic Brain. *Front. Biosci.* **2006**, *11* (1), 1289–1301. https://doi.org/10.2741/1883.

(36) Rivera, S.; Ogier, C.; Jourquin, J.; Timsit, S.; Szklarczyk, A. W.; Miller, K.; Gearing, A. J. H.; Kaczmarek, L.; Khrestchatisky, M. Gelatinase B and TIMP-1 Are Regulated in a Cell- and Time-Dependent Manner in Association with Neuronal Death and Glial Reactivity after Global Forebrain Ischemia. *Eur. J. Neurosci.* **2002**, *15* (1), 19–32. https://doi.org/10.1046/j.0953-816x.2001.01838.x.

(37) Gu, Z. S-Nitrosylation of Matrix Metalloproteinases: Signaling Pathway to Neuronal Cell Death. *Science (5584).* **2002**, *297* (5584), 1186–1190. https://doi.org/10.1126/science.1073634.

(38) Muri, L.; Leppert, D.; Grandgirard, D.; Leib, S. L. MMPs and ADAMs in Neurological Infectious Diseases and Multiple Sclerosis. *Cell. Mol. Life Sci.* **2019**, *76* (16), 3097–3116. https://doi.org/10.1007/s00018-019-03174-6.

(39) Montaner, J.; Ramiro, L.; Simats, A.; Hernández-Guillamon, M.; Delgado, P.; Bustamante, A.; Rosell, A. Matrix Metalloproteinases and ADAMs in Stroke. *Cell. Mol. Life Sci.* **2019**, *76* (16), 3117–3140. https://doi.org/10.1007/s00018-019-03175-5.

(40) Rosenberg, G. A. Matrix Metalloproteinases in Neuroinflammation. *Glia* **2002**, *39* (3), 279–291. https://doi.org/10.1002/glia.10108.

(41) Chopra, S.; Overall, C. M.; Dufour, A. Matrix Metalloproteinases in the CNS: Interferons Get Nervous. *Cell. Mol. Life Sci.* **2019**, *76* (16), 3083–3095. https://doi.org/10.1007/s00018-019-03171-9.

(42) Nakada, M.; Okada, Y.; Yamashita, J. The Role of Matrix Metalloproteinases in Glioma Invasion. *Front. Biosci.* **2003**, *8*, 261–269.

(43) Martins, D.; Moreira, J.; Gonçalves, N. P.; Saraiva, M. J. MMP-14 Overexpression Correlates with the Neurodegenerative Process in Familial Amyloidotic Polyneuropathy. *Dis. Model. Mech.* **2017**, *10* (10), 1253–1260. https://doi.org/10.1242/dmm.028571.

(44) Rosenberg, G. A. Matrix Metalloproteinases and Their Multiple Roles in Neurodegenerative Diseases. *Lancet Neurol.* **2009**, *8* (2), 205–216. https://doi.org/10.1016/S1474-4422(09)70016-X.

(45) Trivedi, A.; Noble-Haeusslein, L. J.; Levine, J. M.; Santucci, A. D.; Reeves, T. M.; Phillips, L. L. Matrix Metalloproteinase Signals Following Neurotrauma Are Right on Cue. *Cell. Mol. Life Sci.* **2019**, *76* (16), 3141–3156. https://doi.org/10.1007/s00018-019-03176-4.

(46) Leake, A.; Morris, C. .; Whateley, J. Brain Matrix Metalloproteinase 1 Levels Are Elevated in Alzheimer's Disease. *Neurosci. Lett.* **2000**, *291* (3), 201–203. https://doi.org/10.1016/S0304-3940(00)01418-X.

(47) Py, N. A.; Bonnet, A. E.; Bernard, A.; Marchalant, Y.; Charrat, E.; Checler, F.; Khrestchatisky, M.; Baranger, K.; Rivera, S. Differential Spatio-Temporal Regulation of MMPs in the 5xFAD Mouse Model of Alzheimer's Disease: Evidence for a pro-Amyloidogenic Role of MT1-MMP. *Front. Aging Neurosci.* **2014**, *6*, 1–17. https://doi.org/10.3389/fnagi.2014.00247.

(48) Yoshiyama, Y.; Asahina, M.; Hattori, T. Selective Distribution of Matrix Metalloproteinase-3 (MMP-3) in Alzheimer's Disease Brain. *Acta Neuropathol.* **2000**, *99* (2), 91–95. https://doi.org/10.1007/PL00007428.

(49) Zhu, B. L.; Long, Y.; Luo, W.; Yan, Z.; Lai, Y. J.; Zhao, L. G.; Zhou, W. H.; Wang, Y. J.; Shen, L. L.; Liu, L.; Deng, X-J.; Wang, X-F.; Sun, F.; Chen, G-J. MMP13 Inhibition Rescues Cognitive Decline in Alzheimer Transgenic Mice via BACE1 Regulation. *Brain* **2019**, *142* (1), 176–192. https://doi.org/10.1093/brain/awy305.

(50) Langenfurth, A.; Rinnenthal, J. L.; Vinnakota, K.; Prinz, V.; Carlo, A.-S.; Stadelmann, C.; Siffrin, V.; Peaschke, S.; Endres, M.; Heppner, F.; Glass, R.; Wolf, S. A.; Kettenmann, H. Membrane-Type 1 Metalloproteinase Is Upregulated in Microglia/brain Macrophages in Neurodegenerative and Neuroinflammatory Diseases. *J. Neurosci. Res.* **2014**, *92* (3), 275–286. https://doi.org/10.1002/jnr.23288.

(51) Dawkins, E.; Small, D. H. Insights into the Physiological Function of the β-Amyloid Precursor Protein: Beyond Alzheimer's Disease. *J. Neurochem.* **2014**, *129* (5), 756–769. https://doi.org/10.1111/jnc.12675.

(52) Müller, U. C.; Deller, T.; Korte, M. Not Just Amyloid: Physiological Functions of the Amyloid Precursor Protein Family. *Nat. Rev. Neurosci.* **2017**, *18* (5), 281–298. https://doi.org/10.1038/nrn.2017.29.

(53) Zhang, Y.; Thompson, R.; Zhang, H.; Xu, H. APP Processing in Alzheimer's Disease. *Mol. Brain* **2011**, *4* (1), 3. https://doi.org/10.1186/1756-6606-4-3.

(54) Coronel, R.; Bernabeu-Zornoza, A.; Palmer, C.; Muñiz-Moreno, M.; Zambrano, A.; Cano, E.; Liste, I. Role of Amyloid Precursor Protein (APP) and Its Derivatives in the Biology and Cell Fate Specification of Neural Stem Cells. *Mol. Neurobiol.* **2018**, *55* (9), 7107–7117. https://doi.org/10.1007/s12035-018-0914-2.

(55) Lauritzen, I.; Pardossi-Piquard, R.; Bauer, C.; Brigham, E.; Abraham, J.-D.; Ranaldi, S.; Fraser, P.; St-George-Hyslop, P.; Le Thuc, O.; Espin, V.; Chami, L.; Dunys, J.; Checler, F. The β-Secretase-Derived C-Terminal Fragment of APP, C99, But Not A , Is a Key Contributor to Early Intraneuronal Lesions in Triple-Transgenic Mouse Hippocampus. *J. Neurosci.* **2012**, *32* (46), 16243–16255. https://doi.org/10.1523/JNEUROSCI.2775-12.2012.

(56) Bukhari, H.; Glotzbach, A.; Kolbe, K.; Leonhardt, G.; Loosse, C.; Müller, T. Small Things Matter: Implications of APP Intracellular Domain AICD Nuclear Signaling in the Progression and Pathogenesis of Alzheimer's Disease. *Prog. Neurobiol.* **2017**, *156*, 189–213. https://doi.org/10.1016/j.pneurobio.2017.05.005.

(57) Carson, J. A.; Turner, A. J. β-Amyloid Catabolism: Roles for Neprilysin (NEP) and Other Metallopeptidases? *J. Neurochem.* **2002**, *81* (1), 1–8. https://doi.org/10.1046/j.1471-4159.2002.00855.x.

(58) O'Brien, R. J.; Wong, P. C. Amyloid Precursor Protein Processing and Alzheimer's Disease. *Annu. Rev. Neurosci.* **2011**, *34* (1), 185–204. https://doi.org/10.1146/annurev-neuro-061010-113613.

(59) Roher, A. E.; Kasunic, T. C.; Woods, A. S.; Cotter, R. J.; Ball, M. J.; Fridman, R. Proteolysis of Aβ Peptide from Alzheimer Disease Brain by Gelatinase A. *Biochem. Biophys. Res. Commun.* **1994**, *205* (3), 1755–1761. https://doi.org/10.1006/bbrc.1994.2872.

(60) Backstrom, J. R.; Lim, G. P.; Cullen, M. J.; Tökés, Z. A. Matrix Metalloproteinase-9

(MMP-9) Is Synthesized in Neurons of the Human Hippocampus and Is Capable of Degrading the Amyloid-β Peptide (1–40). *J. Neurosci.* **1996**, *16* (24), 7910–7919. https://doi.org/10.1523/JNEUROSCI.16-24-07910.1996.

(61) Yan, P.; Hu, X.; Song, H.; Yin, K.; Bateman, R. J.; Cirrito, J. R.; Xiao, Q.; Hsu, F. F.; Turk, J. W.; Xu, J.; Hsu, C. Y.; Holtzman, D. M.; Lee, J-M. Matrix Metalloproteinase-9 Degrades Amyloid-β Fibrils in Vitro and Compact Plaques in Situ. *J. Biol. Chem.* **2006**, *281* (34), 24566–24574. https://doi.org/10.1074/jbc.M602440200.

(62) Hernandez-Guillamon, M.; Mawhirt, S.; Blais, S.; Montaner, J.; Neubert, T. A.; Rostagno, A.; Ghiso, J. Sequential Amyloid-β Degradation by the Matrix Metalloproteases MMP-2 and MMP-9. *J. Biol. Chem.* **2015**, *290* (24), 15078–15091. https://doi.org/10.1074/jbc.M114.610931.

(63) Yin, K.-J.; Cirrito, J. R.; Yan, P.; Hu, X.; Xiao, Q.; Pan, X.; Bateman, R.; Song, H.; Hsu, F.-F.; Turk, J.; Xu, J.; Hsu, C. Y.; Mills, J. C.; Holtzman, D. M.; Lee, J-M. Matrix Metalloproteinases Expressed by Astrocytes Mediate Extracellular Amyloid-Beta Peptide Catabolism. *J. Neurosci.* **2006**, *26* (43), 10939–10948. https://doi.org/10.1523/JNEUROSCI.2085-06.2006.

(64) Koronyo-Hamaoui, M.; Ko, M. K.; Koronyo, Y.; Azoulay, D.; Seksenyan, A.; Kunis, G.; Pham, M.; Bakhsheshian, J.; Rogeri, P.; Black, K. L.; Farkas, D. L.; Schwartz, M. Attenuation of AD-like Neuropathology by Harnessing Peripheral Immune Cells: Local Elevation of IL-10 and MMP-9. *J. Neurochem.* **2009**, *111* (6), 1409–1424. https://doi.org/10.1111/j.1471-4159.2009.06402.x.

(65) Fragkouli, A.; Tsilibary, E. C.; Tzinia, A. K. Neuroprotective Role of MMP-9 Overexpression in the Brain of Alzheimer's 5xFAD Mice. *Neurobiol. Dis.* **2014**, *70*, 179–189. https://doi.org/10.1016/j.nbd.2014.06.021.

(66) Bruno, M. A.; Mufson, E. J.; Wuu, J.; Cuello, A. C. Increased Matrix Metalloproteinase 9 Activity in Mild Cognitive Impairment. *J. Neuropathol. Exp. Neurol.* **2009**, *68* (12), 1309–1318. https://doi.org/10.1097/NEN.0b013e3181c22569.

(67) Wilcock, D. M.; Morgan, D.; Gordon, M. N.; Taylor, T. L.; Ridnour, L. A.; Wink, D. A.; Colton, C. A. Activation of Matrix Metalloproteinases Following Anti-Aβ Immunotherapy; Implications for Microhemorrhage Occurrence. *J. Neuroinflammation* **2011**, *8* (1), 1–13. https://doi.org/10.1186/1742-2094-8-115.

(68) Bell, R. D.; Winkler, E. A.; Singh, I.; Sagare, A. P.; Deane, R.; Wu, Z.; Holtzman, D. M.; Betsholtz, C.; Armulik, A.; Sallstrom, J.; Berk, B. C.; Zlokovic, B. V. Apolipoprotein E Controls Cerebrovascular Integrity via Cyclophilin A. *Nature* **2012**, *485* (7399), 512–516. https://doi.org/10.1038/nature11087.

(69) Liao, M. C.; Van Nostrand, W. E. Degradation of Soluble and Fibrillar Amyloid β-Protein by Matrix Metalloproteinase (MT1-MMP) in Vitro. *Biochemistry* **2010**, *49* (6), 1127–1136. https://doi.org/10.1021/bi901994d.

(70) Li, W.; Poteet, E.; Xie, L.; Liu, R.; Wen, Y.; Yang, S.-H. Regulation of Matrix Metalloproteinase 2 by Oligomeric Amyloid β Protein. *Brain Res.* **2011**, *1387*, 141–148. https://doi.org/10.1016/j.brainres.2011.02.078.

(71) Shlosberg, D.; Benifla, M.; Kaufer, D.; Friedman, A. Blood–brain Barrier Breakdown as a

Therapeutic Target in Traumatic Brain Injury. *Nat. Rev. Neurol.* **2010**, *6* (7), 393–403. https://doi.org/10.1038/nrneurol.2010.74.

(72) Sweeney, M. D.; Sagare, A. P.; Zlokovic, B. V. Blood–brain Barrier Breakdown in Alzheimer Disease and Other Neurodegenerative Disorders. *Nat. Rev. Neurol.* **2018**, *14* (3), 133–150. https://doi.org/10.1038/nrneurol.2017.188.

(73) Giannoni, P.; Arango-Lievano, M.; Neves, I. Das; Rousset, M.-C.; Baranger, K.; Rivera, S.; Jeanneteau, F.; Claeysen, S.; Marchi, N. Cerebrovascular Pathology during the Progression of Experimental Alzheimer's Disease. *Neurobiol. Dis.* **2016**, *88*, 107–117. https://doi.org/10.1016/j.nbd.2016.01.001.

(74) Terni, B.; Ferrer, I. Abnormal Expression and Distribution of MMP2 at Initial Stages of Alzheimer's Disease-Related Pathology. *J. Alzheimer's Dis.* **2015**, *46* (2), 461–469. https://doi.org/10.3233/JAD-142460.

(75) Asahina, M.; Yoshiyama, Y.; Hattori, T. Expression of Matrix Metalloproteinase-9 and Urinary-Type Plasminogen Activator in Alzheimer's Disease Brain. *Clin. Neuropathol.* **2001**, *20* (2), 60–63.

(76) Holmes, C. Systemic Infection, Interleukin 1beta, and Cognitive Decline in Alzheimer's Disease. *J. Neurol. Neurosurg. Psychiatry* **2003**, *74* (6), 788–789. https://doi.org/10.1136/jnnp.74.6.788.

(77) Paumier, J.-M.; Py, N. A.; García-González, L.; Bernard, A.; Stephan, D.; Louis, L.; Checler, F.; Khrestchatisky, M.; Baranger, K.; Rivera, S. Proamyloidogenic Effects of Membrane Type 1 Matrix Metalloproteinase Involve MMP-2 and BACE-1 Activities, and the Modulation of APP Trafficking. *FASEB J.* **2019**, *33* (2), 2910–2927. https://doi.org/10.1096/fj.201801076R.

(78) Holmbeck, K.; Bianco, P.; Caterina, J.; Yamada, S.; Kromer, M.; Kuznetsov, S. A.; Mankani, M.; Gehron Robey, P.; Poole, A. R.; Pidoux, I.; Ward, J. M.; Birkedal-Hansen, H. MT1-MMP-Deficient Mice Develop Dwarfism, Osteopenia, Arthritis, and Connective Tissue Disease due to Inadequate Collagen Turnover. *Cell* **1999**, *99* (1), 81–92. https://doi.org/10.1016/S0092-8674(00)80064-1.

(79) Zhou, Z.; Apte, S. S.; Soininen, R.; Cao, R.; Baaklini, G. Y.; Rauser, R. W.; Wang, J.; Cao, Y.; Tryggvason, K. Impaired Endochondral Ossification and Angiogenesis in Mice Deficient in Membrane-Type Matrix Metalloproteinase I. *Proc. Natl. Acad. Sci.* **2000**, *97* (8), 4052–4057. https://doi.org/10.1073/pnas.060037197.

(80) Remacle, A. G.; Cieplak, P.; Nam, D. H.; Shiryaev, S. A.; Ge, X.; Strongin, A. Y. Selective Function-Blocking Monoclonal Human Antibody Highlights the Important Role of Membrane Type-1 Matrix Metalloproteinase (MT1-MMP) in Metastasis. *Oncotarget* **2017**, *8* (2), 2781–2799. https://doi.org/10.18632/oncotarget.13157.

(81) Udi, Y.; Grossman, M.; Solomonov, I.; Dym, O.; Rozenberg, H.; Moreno, V.; Cuniasse, P.; Dive, V.; Arroyo, A. G.; Sagi, I. Inhibition Mechanism of Membrane Metalloprotease by an Exosite-Swiveling Conformational Antibody. *Structure* **2015**, *23* (1), 104–115. https://doi.org/10.1016/j.str.2014.10.012.

(82) Miyazaki, K.; Hasegawa, M.; Funahashi, K.; Umeda, M. A Metalloproteinase Inhibitor Domain in Alzheimer Amyloid Protein Precursor. *Nature* **1993**, *362* (6423), 839–841.

https://doi.org/10.1038/362839a0.

(83) Higashi, S.; Miyazaki, K. Identification of a Region of β-Amyloid Precursor Protein Essential for Its Gelatinase A Inhibitory Activity. *J. Biol. Chem.* **2003**, *278* (16), 14020–14028. https://doi.org/10.1074/jbc.M212264200.

(84) Higashi, S.; Miyazaki, K. Identification of Amino Acid Residues of the Matrix Metalloproteinase-2 Essential for Its Selective Inhibition by β-Amyloid Precursor Protein-Derived Inhibitor. *J. Biol. Chem.* **2008**, *283* (15), 10068–10078. https://doi.org/10.1074/jbc.M709509200.

(85) Hashimoto, H.; Takeuchi, T.; Komatsu, K.; Miyazaki, K.; Sato, M.; Higashi, S. Structural Basis for Matrix Metalloproteinase-2 (MMP-2)-Selective Inhibitory Action of β-Amyloid Precursor Protein-Derived Inhibitor. *J. Biol. Chem.* **2011**, *286* (38), 33236–33243. https://doi.org/10.1074/jbc.M111.264176.

(86) Higashi, S.; Hirose, T.; Takeuchi, T.; Miyazaki, K. Molecular Design of a Highly Selective and Strong Protein Inhibitor against Matrix Metalloproteinase-2 (MMP-2). *J. Biol. Chem.* **2013**, *288* (13), 9066–9076. https://doi.org/10.1074/jbc.M112.441758.

(87) Higashi, S.; Miyazaki, K. Novel Processing of β-Amyloid Precursor Protein Catalyzed by Membrane Type 1 Matrix Metalloproteinase Releases a Fragment Lacking the Inhibitor Domain against Gelatinase A †. *Biochemistry* **2003**, *42* (21), 6514–6526. https://doi.org/10.1021/bi020643m.

(88) Sato, H.; Takino, T.; Kinoshita, T.; Imai, K.; Okada, Y.; Stetler Stevenson, W. G.; Seiki, M. Cell Surface Binding and Activation of Gelatinase A Induced by Expression of Membrane-Type-1-Matrix Metalloproteinase (MT1-MMP). *FEBS Lett.* **1996**, *385* (3), 238–240. https://doi.org/10.1016/0014-5793(96)00389-4.

(89) Haapasalo, A.; Kovacs, D. M. The Many Substrates of Presenilin/γ-Secretase. *J. Alzheimer's Dis.* **2011**, *25* (1), 3–28. https://doi.org/10.3233/JAD-2011-101065.

(90) Paumier, J. M.; Thinakaran, G. Matrix Metalloproteinase 13, a New Target for Therapy in Alzheimer's Disease. *Genes Dis.* **2019**, *6* (1), 1–2. https://doi.org/10.1016/j.gendis.2019.02.004.

(91) Ito, S.; Kimura, K.; Haneda, M.; Ishida, Y.; Sawada, M.; Isobe, K. Induction of Matrix Metalloproteinases (MMP3, MMP12 and MMP13) Expression in the Microglia by Amyloid-β Stimulation via the PI3K/Akt Pathway. *Exp. Gerontol.* **2007**, *42* (6), 532–537. https://doi.org/10.1016/j.exger.2006.11.012.

(92) Walker, D. G.; Link, J.; Lue, L.; Dalsing-Hernandez, J. E.; Boyes, B. E. Gene Expression Changes by Amyloid β Peptide-Stimulated Human Postmortem Brain Microglia Identify Activation of Multiple Inflammatory Processes. *J. Leukoc. Biol.* **2006**, *79* (3), 596–610. https://doi.org/10.1189/jlb.0705377.

(93) Ierusalimsky, V. N.; Kuleshova, E. P.; Balaban, P. M. Expression of the Type 1 Metalloproteinase in the Rat Hippocampus after the Intracerebroventricular Injection of β-Amyloid Peptide (25-35). *Acta Neurobiol. Exp. (Wars).* **2013**, *73* (4), 571–578.

(94) Lin, J.; Kakkar, V.; Lu, X. Impact of Matrix Metalloproteinases on Atherosclerosis. *Curr. Drug Targets* **2014**, *15* (4), 442–453. https://doi.org/10.2174/1389450115666140211115805.

(95) Johnson, J. L. Metalloproteinases in Atherosclerosis. *Eur. J. Pharmacol.* **2017**, *816* (6), 93–106. https://doi.org/10.1016/j.ejphar.2017.09.007.

(96) Elkington, P.; Shiomi, T.; Breen, R.; Nuttall, R. K.; Ugarte-Gil, C. A.; Walker, N. F.; Saraiva, L.; Pedersen, B.; Mauri, F.; Lipman, M.; Edwards, D. R.; Robertson, B. D.; D'Armiento, J.; Friedland, J. S. MMP-1 Drives Immunopathology in Human Tuberculosis and Transgenic Mice. *J. Clin. Invest.* **2011**, *121* (5), 1827–1833. https://doi.org/10.1172/JCI45666.

(97) Chen, Q.; Jin, M.; Yang, F.; Zhu, J.; Xiao, Q.; Zhang, L. Matrix Metalloproteinases: Inflammatory Regulators of Cell Behaviors in Vascular Formation and Remodeling. *Mediators Inflamm.* **2013**, *2013*, 1–14. https://doi.org/10.1155/2013/928315.

(98) Barichello, T.; Generoso, J. S.; Michelon, C. M.; Simões, L. R.; Elias, S. G.; Vuolo, F.; Comim, C. M.; Dal-Pizzol, F.; Quevedo, J. Inhibition of Matrix Metalloproteinases-2 and -9 Prevents Cognitive Impairment Induced by Pneumococcal Meningitis in Wistar Rats. *Exp. Biol. Med.* **2014**, *239* (2), 225–231. https://doi.org/10.1177/1535370213508354.

(99) Hannocks, M.-J.; Zhang, X.; Gerwien, H.; Chashchina, A.; Burmeister, M.; Korpos, E.; Song, J.; Sorokin, L. The Gelatinases, MMP-2 and MMP-9, as Fine Tuners of Neuroinflammatory Processes. *Matrix Biol.* **2019**, *75–76*, 102–113. https://doi.org/10.1016/j.matbio.2017.11.007.

(100) White, A. R.; Du, T.; Laughton, K. M.; Volitakis, I.; Sharples, R. A.; Xilinas, M. E.; Hoke, D. E.; Holsinger, R. M. D.; Evin, G.; Cherny, R. A.; Hill, A. F.; Barnham, K. J.; Li, Q-X.; Bush, A. I.; Masters, C. L. Degradation of the Alzheimer Disease Amyloid β-Peptide by Metal-Dependent Up-Regulation of Metalloprotease Activity. *J. Biol. Chem.* **2006**, *281* (26), 17670–17680. https://doi.org/10.1074/jbc.M602487200.

(101) Brkic, M.; Balusu, S.; Van Wonterghem, E.; Gorle, N.; Benilova, I.; Kremer, A.; Van Hove, I.; Moons, L.; De Strooper, B.; Kanazir, S.; Libert, C.; Vandenbroucke, R. E. Amyloid Oligomers Disrupt Blood-CSF Barrier Integrity by Activating Matrix Metalloproteinases. *J. Neurosci.* **2015**, *35* (37), 12766–12778. https://doi.org/10.1523/JNEUROSCI.0006-15.2015.

(102) Kim, H. J.; Fillmore, H. L.; Reeves, T. M.; Phillips, L. L. Elevation of Hippocampal MMP-3 Expression and Activity during Trauma-Induced Synaptogenesis. *Exp. Neurol.* **2005**, *192* (1), 60–72. https://doi.org/10.1016/j.expneurol.2004.10.014.

(103) Choi, D. H.; Kim, Y. J.; Kim, Y. G.; Joh, T. H.; Beal, M. F.; Kim, Y. S. Role of Matrix Metalloproteinase 3-Mediated α-Synuclein Cleavage in Dopaminergic Cell Death. *J. Biol. Chem.* **2011**, *286* (16), 14168–14177. https://doi.org/10.1074/jbc.M111.222430.

(104) Chung, Y. C.; Kim, Y.-S.; Bok, E.; Yune, T. Y.; Maeng, S.; Jin, B. K. MMP-3 Contributes to Nigrostriatal Dopaminergic Neuronal Loss, BBB Damage, and Neuroinflammation in an MPTP Mouse Model of Parkinson's Disease. *Mediators Inflamm.* **2013**, 1–11. https://doi.org/10.1155/2013/370526.

(105) Stawarski, M.; Stefaniuk, M.; Wlodarczyk, J. Matrix Metalloproteinase-9 Involvement in the Structural Plasticity of Dendritic Spines. *Front. Neuroanat.* **2014**, *8* (7), 1–15. https://doi.org/10.3389/fnana.2014.00068.

(106) Murase, S.; Lantz, C. L.; Kim, E.; Gupta, N.; Higgins, R.; Stopfer, M.; Hoffman, D. A.;

Quinlan, E. M. Matrix Metalloproteinase-9 Regulates Neuronal Circuit Development and Excitability. *Mol. Neurobiol.* **2016**, *53* (5), 3477–3493. https://doi.org/10.1007/s12035-015-9295-y.

(107) Larsen, P. H.; DaSilva, A. G.; Conant, K.; Yong, V. W. Myelin Formation during Development of the CNS Is Delayed in Matrix Metalloproteinase-9 and -12 Null Mice. *J. Neurosci.* **2006**, *26* (8), 2207–2214. https://doi.org/10.1523/JNEUROSCI.1880-05.2006.

(108) Liu, Y.; Zhang, M.; Hao, W.; Mihaljevic, I.; Liu, X.; Xie, K.; Walter, S.; Fassbender, K. Matrix Metalloproteinase-12 Contributes to Neuroinflammation in the Aged Brain. *Neurobiol. Aging* **2013**, *34* (4), 1231–1239. https://doi.org/10.1016/j.neurobiolaging.2012.10.015.

(109) Liao, G.; Wang, Z.; Lee, E.; Moreno, S.; Abuelnasr, O.; Baudry, M.; Bi, X. Enhanced Expression of Matrix Metalloproteinase-12 Contributes to Npc1 Deficiency-Induced Axonal Degeneration. *Exp. Neurol.* **2015**, *269*, 67–74. https://doi.org/10.1016/j.expneurol.2015.04.004.

(110) Goncalves DaSilva, A.; Yong, V. W. Matrix Metalloproteinase-12 Deficiency Worsens Relapsing-Remitting Experimental Autoimmune Encephalomyelitis in Association with Cytokine and Chemokine Dysregulation. *Am. J. Pathol.* **2009**, *174* (3), 898–909. https://doi.org/10.2353/ajpath.2009.080952.

(111) Goncalves DaSilva, A.; Liaw, L.; Yong, V. W. Cleavage of Osteopontin by Matrix Metalloproteinase-12 Modulates Experimental Autoimmune Encephalomyelitis Disease in C57BL/6 Mice. *Am. J. Pathol.* **2010**, *177* (3), 1448–1458.

https://doi.org/10.2353/ajpath.2010.091081.

(112) Dufour, A.; Bellac, C. L.; Eckhard, U.; Solis, N.; Klein, T.; Kappelhoff, R.; Fortelny, N.; Jobin, P.; Rozmus, J.; Mark, J.; Pavlidis, P.; Dive, V.; Barbour, S. J.; Overall, C. M. C-Terminal Truncation of IFN-γ Inhibits Proinflammatory Macrophage Responses and Is Deficient in Autoimmune Disease. *Nat. Commun.* **2018**, *9* (1), 1–18. https://doi.org/10.1038/s41467-018-04717-4.

(113) Bellac, C. L.; Dufour, A.; Krisinger, M. J.; Loonchanta, A.; Starr, A. E.; auf dem Keller, U.; Lange, P. F.; Goebeler, V.; Kappelhoff, R.; Butler, G. S.; Burtnick, L. D.; Conway, E. M.; Roberts, C. R.; Overall, C. M. Macrophage Matrix Metalloproteinase-12 Dampens Inflammation and Neutrophil Influx in Arthritis. *Cell Rep.* **2014**, *9* (2), 618–632. https://doi.org/10.1016/j.celrep.2014.09.006.

(114) Houghton, A. M.; Hartzell, W. O.; Robbins, C. S.; Gomis-Rüth, F. X.; Shapiro, S. D. Macrophage Elastase Kills Bacteria within Murine Macrophages. *Nature* **2009**, *460* (7255), 637–641. https://doi.org/10.1038/nature08181.

(115) Marchant, D. J.; Bellac, C. L.; Moraes, T. J.; Wadsworth, S. J.; Dufour, A.; Butler, G. S.; Bilawchuk, L. M.; Hendry, R. G.; Robertson, A. G.; Cheung, C. T.; Ng, J.; Ang, L.; Luo, Z.; Heilbron, K.; Norris, M. J.; Duan, W.; Bucyk, T.; Karpov, A.; Devel, L.; Georgiadis, D.; Hegele, R. G.; Luo, H.; Granville, D. J.; Dive, V.; McManus, B. M.; Overall, C. M. A New Transcriptional Role for Matrix Metalloproteinase-12 in Antiviral Immunity. *Nat. Med.* **2014**, *20* (5), 493–502. https://doi.org/10.1038/nm.3508.

(116) Houghton, A. M.; Grisolano, J. L.; Baumann, M. L.; Kobayashi, D. K.; Hautamaki, R. D.;

Nehring, L. C.; Cornelius, L. A.; Shapiro, S. D. Macrophage Elastase (Matrix Metalloproteinase-12) Suppresses Growth of Lung Metastases. *Cancer Res.* **2006**, *66* (12), 6149–6155. https://doi.org/10.1158/0008-5472.CAN-04-0297.

(117) Dandachi, N.; Kelly, N. J.; Wood, J. P.; Burton, C. L.; Radder, J. E.; Leme, A. S.; Gregory, A. D.; Shapiro, S. D. Macrophage Elastase Induces TRAIL-Mediated Tumor Cell Death through Its Carboxy-Terminal Domain. *Am. J. Respir. Crit. Care Med.* **2017**, *196* (3), 353–363. https://doi.org/10.1164/rccm.201606-1150OC.

(118) Ma, F.; Martínez-San Segundo, P.; Barceló, V.; Morancho, A.; Gabriel-Salazar, M.; Giralt, D.; Montaner, J.; Rosell, A. Matrix Metalloproteinase-13 Participates in Neuroprotection and Neurorepair after Cerebral Ischemia in Mice. *Neurobiol. Dis.* **2016**, *91*, 236–246. https://doi.org/10.1016/j.nbd.2016.03.016.

(119) Ahmad, M.; Takino, T.; Miyamori, H.; Yoshizaki, T.; Furukawa, M.; Sato, H. Cleavage of Amyloid-β Precursor Protein (APP) by Membrane-Type Matrix Metalloproteinases. *J. Biochem.* **2006**, *139* (3), 517–526. https://doi.org/10.1093/jb/mvj054.

(120) Ulasov, I.; Yi, R.; Guo, D.; Sarvaiya, P.; Cobbs, C. The Emerging Role of MMP14 in Brain Tumorigenesis and Future Therapeutics. *Biochim. Biophys. Acta - Rev. Cancer* **2014**, *1846* (1), 113–120. https://doi.org/10.1016/j.bbcan.2014.03.002.

(121) Sakamoto, T.; Seiki, M. Integrated Functions of Membrane-Type 1 Matrix Metalloproteinase in Regulating Cancer Malignancy: Beyond a Proteinase. *Cancer Sci.* **2017**, *108* (6), 1095–1100. https://doi.org/10.1111/cas.13231.

(122) Talmi-Frank, D.; Altboum, Z.; Solomonov, I.; Udi, Y.; Jaitin, D. A.; Klepfish, M.; David,

E.; Zhuravlev, A.; Keren-Shaul, H.; Winter, D. R.; Gat-Viks, I.; Mandelboim, M.; Ziv, T.; Amit, I.; Sagi, I. Extracellular Matrix Proteolysis by MT1-MMP Contributes to Influenza-Related Tissue Damage and Mortality. *Cell Host Microbe* **2016**, *20* (4), 458–470. https://doi.org/10.1016/j.chom.2016.09.005.

(123) Aguirre, A.; Blázquez-Prieto, J.; Amado-Rodriguez, L.; López-Alonso, I.; Batalla-Solís, E.; González-López, A.; Sánchez-Pérez, M.; Mayoral-Garcia, C.; Gutiérrez-Fernández, A.; Albaiceta, G. M. Matrix Metalloproteinase-14 Triggers an Anti-Inflammatory Proteolytic Cascade in Endotoxemia. *J. Mol. Med.* **2017**, *95* (5), 487–497. https://doi.org/10.1007/s00109-017-1510-z.

(124) Sakamoto, T.; Seiki, M. Cytoplasmic Tail of MT1-MMP Regulates Macrophage Motility Independently from Its Protease Activity. *Genes to Cells* **2009**, *14* (5), 617–626. https://doi.org/10.1111/j.1365-2443.2009.01293.x.

(125) Baranger, K.; Marchalant, Y.; Bonnet, A. E.; Crouzin, N.; Carrete, A.; Paumier, J.-M.; Py, N. A.; Bernard, A.; Bauer, C.; Charrat, E.; Moschke, K.; Seiki, M.; Vignes, M.; Lichtenthaler, S. F.; Checler, F.; Khrestchatisky, M.; Rivera, S. MT5-MMP Is a New pro-Amyloidogenic Proteinase That Promotes Amyloid Pathology and Cognitive Decline in a Transgenic Mouse Model of Alzheimer's Disease. *Cell. Mol. Life Sci.* **2016**, *73* (1), 217–236. https://doi.org/10.1007/s00018-015-1992-1.

(126) Baranger, K.; Bonnet, A. E.; Girard, S. D.; Paumier, J.-M.; García-González, L.; Elmanaa, W.; Bernard, A.; Charrat, E.; Stephan, D.; Bauer, C.; Moschke, K.; Lichtenthaler, S. F.; Roman, F. S.; Checler, F.; Khrestchatisky, M.; Rivera, S. MT5-MMP Promotes Alzheimer's Pathogenesis in the Frontal Cortex of 5xFAD Mice and APP Trafficking in Vitro. *Front. Mol. Neurosci.* **2017**, *9* (1), 1–17. https://doi.org/10.3389/fnmol.2016.00163.

(127) Baranger, K.; Khrestchatisky, M.; Rivera, S. MT5-MMP, Just a New APP Processing Proteinase in Alzheimer's Disease? *J. Neuroinflammation* **2016**, *13* (1), 167. https://doi.org/10.1186/s12974-016-0633-4.

(128) Willem, M.; Tahirovic, S.; Busche, M. A.; Ovsepian, S. V.; Chafai, M.; Kootar, S.; Hornburg, D.; Evans, L. D. B.; Moore, S.; Daria, A.; Hampel, H.; Müller, V.; Giudici, C.; Nuscher, B.; Wenninger-Weinzierl, A.; Kremmer, E.; Heneka, M. T.; Thal, D. R.; Giedraitis, V.; Lannfelt, L.; Müller, U.; Livesey, F. J.; Meissner, F; Herms, J.; Konnerth, A.; Marie, H.; Haass, C. η-Secretase Processing of APP Inhibits Neuronal Activity in the Hippocampus. *Nature* **2015**, *526* (7573), 443–447. https://doi.org/10.1038/nature14864.

(129) Hayashita-Kinoh, H.; Kinoh, H.; Okada, a; Komori, K.; Itoh, Y.; Chiba, T.; Kajita, M.; Yana, I.; Seiki, M. Membrane-Type 5 Matrix Metalloproteinase Is Expressed in Differentiated Neurons and Regulates Axonal Growth. *Cell Growth Differ.* **2001**, *12* (11), 573–580.

(130) Komori, K.; Nonaka, T.; Okada, A.; Kinoh, H.; Hayashita-Kinoh, H.; Yoshida, N.; Yana, I.; Seiki, M. Absence of Mechanical Allodynia and Aβ-Fiber Sprouting after Sciatic Nerve Injury in Mice Lacking Membrane-Type 5 Matrix Metalloproteinase. *FEBS Lett.* **2004**, *557* (1–3), 125–128. https://doi.org/10.1016/S0014-5793(03)01458-3.

(131) Folgueras, A. R.; Valdes-Sanchez, T.; Llano, E.; Menendez, L.; Baamonde, A.; Denlinger, B. L.; Belmonte, C.; Juarez, L.; Lastra, A.; Garcia-Suarez, O.; Astudillo, A.; Kirstein, M.; Pendas, A. M.; Farinas, I.; Lopez-Otin, C. Metalloproteinase MT5-MMP Is an Essential

Modulator of Neuro-Immune Interactions in Thermal Pain Stimulation. *Proc. Natl. Acad. Sci.* **2009**, *106* (38), 16451–16456. https://doi.org/10.1073/pnas.0908507106.

(132) Okimoto, R. A.; Breitenbuecher, F.; Olivas, V. R.; Wu, W.; Gini, B.; Hofree, M.; Asthana, S.; Hrustanovic, G.; Flanagan, J.; Tulpule, A.; Blakely, C. M.; Haringsma, H. J.; Simmons, A. D.; Gowen, K.; Suh, J.; Miller, V. A.; Ali, S.; Schuler, M.; Bivona, T. G. . Inactivation of Capicua Drives Cancer Metastasis. *Nat. Genet.* **2017**, *49* (1), 87–96. https://doi.org/10.1038/ng.3728.

(133) Llano, E.; Pendás, A. M.; Freije, J. P.; Nakano, A.; Knäuper, V.; Murphy, G.; López-Otin, C. Identification and Characterization of Human MT5-MMP, a New Membrane-Bound Activator of Progelatinase A Overexpressed in Brain Tumors. *Cancer Res.* **1999**, *59*, 2570–2576. https://doi.org/10363975.

(134) Restituito, S.; Khatri, L.; Ninan, I.; Mathews, P. M.; Liu, X.; Weinberg, R. J.; Ziff, E. B. Synaptic Autoregulation by Metalloproteases and -Secretase. *J. Neurosci.* **2011**, *31* (34), 12083–12093. https://doi.org/10.1523/JNEUROSCI.2513-11.2011.

(135) Pei, D. Identification and Characterization of the Fifth Membrane-Type Matrix Metalloproteinase MT5-MMP. *J. Biol. Chem.* **1999**, *274* (13), 8925–8932. https://doi.org/10.1074/jbc.274.13.8925.

(136) Jaworski, D. M. Developmental Regulation of Membrane Type-5 Matrix Metalloproteinase (MT5-MMP) Expression in the Rat Nervous System. *Brain Res.* **2000**, *860* (1–2), 174–177. https://doi.org/10.1016/S0006-8993(00)02035-7.

(137) Monea, S.; Jordan, B. A.; Srivastava, S.; DeSouza, S.; Ziff, E. B. Membrane Localization

of Membrane Type 5 Matrix Metalloproteinase by AMPA Receptor Binding Protein and Cleavage of Cadherins. *J. Neurosci.* **2006**, *26* (8), 2300–2312. https://doi.org/10.1523/JNEUROSCI.3521-05.2006.

(138) García-González, L.; Pilat, D.; Baranger, K.; Rivera, S. Emerging Alternative Proteinases in APP Metabolism and Alzheimer's Disease Pathogenesis: A Focus on MT1-MMP and MT5-MMP. *Front. Aging Neurosci.* **2019**, *11* (1), 1–19. https://doi.org/10.3389/fnagi.2019.00244.

(139) Sekine-Aizawa, Y.; Hama, E.; Watanabe, K.; Tsubuki, S.; Kanai-Azuma, M.; Kanai, Y.; Arai, H.; Aizawa, H.; Iwata, N.; Saido, T. C. Matrix Metalloproteinase (MMP) System in Brain: Identification and Characterization of Brain-Specific MMP Highly Expressed in Cerebellum. *Eur. J. Neurosci.* **2001**, *13* (5), 935–948. https://doi.org/10.1046/j.0953-816x.2001.01462.x.

(140) Andrew, R. J.; Kellett, K. A. B.; Thinakaran, G.; Hooper, N. M. A Greek Tragedy: The Growing Complexity of Alzheimer Amyloid Precursor Protein Proteolysis. *J. Biol. Chem.* **2016**, *291* (37), 19235–19244. https://doi.org/10.1074/jbc.R116.746032.

(141) Nagase, H. Substrate Specificity of MMPs. In *Matrix Metalloproteinase Inhibitors in Cancer Therapy. Cancer Drug Discovery and Development*; Clendeninn N.J., Appelt K., Eds.; Humana Press: Totowa, NJ, 2001; Chapter 2, pp 39–66.

(142) Rodríguez, D.; Morrison, C. J.; Overall, C. M. Matrix Metalloproteinases: What Do They Not Do? New Substrates and Biological Roles Identified by Murine Models and Proteomics. *Biochim. Biophys. Acta - Mol. Cell Res.* **2010**, *1803* (1), 39–54.

https://doi.org/10.1016/j.bbamcr.2009.09.015.

(143) Butler, G. S.; Overall, C. M. Updated Biological Roles for Matrix Metalloproteinases and New "intracellular" substrates Revealed by Degradomics. *Biochemistry* **2009**, *48* (46), 10830–10845. https://doi.org/10.1021/bi901656f.

(144) Maskos, K.; Bode, W. Structural Basis of Matrix Metalloproteinases and Tissue Inhibitors of Metalloproteinases. *Mol. Biotechnol.* **2003**, *25* (3), 241–266. https://doi.org/10.1385/MB:25:3:241.

(145) Bode, W.; Gomis-Rüth, F. X.; Stöckler, W. Astacins, Serralysins, Snake Venom and Matrix Metalloproteinases Exhibit Identical Zinc-Binding Environments (HEXXHXXGXXH and Met-Turn) and Topologies and Should Be Grouped into a Common Family, the "Metzincins." *FEBS Lett.* **1993**, *331* (1–2), 134–140. https://doi.org/10.1016/0014-5793(93)80312-I.

(146) Tallant, C.; Marrero, A.; Gomis-Rüth, F. X. Matrix Metalloproteinases: Fold and Function of Their Catalytic Domains. *Biochim. Biophys. Acta - Mol. Cell Res.* **2010**, *1803* (1), 20–28. https://doi.org/10.1016/j.bbamcr.2009.04.003.

(147) Aureli, L.; Gioia, M.; Cerbara, I.; Monaco, S.; Fasciglione, G.; Marini, S.; Ascenzi, P.; Topai, A.; Coletta, M. Structural Bases for Substrate and Inhibitor Recognition by Matrix Metalloproteinases. *Curr. Med. Chem.* **2008**, *15* (22), 2192–2222. https://doi.org/10.2174/092986708785747490.

(148) Schechter, I.; Berger, A. On the Size of the Active Site in Proteases. I. Papain. *Biochem. Biophys. Res. Commun.* **1967**, *27* (2), 157–162. https://doi.org/10.1016/S0006-

291X(67)80055-X.

(149) The PyMOL Molecular Graphics System, Version 2.3.2, Schrödinger, LLC.

(150) Fabre, B.; Ramos, A.; De Pascual-Teresa, B. Targeting Matrix Metalloproteinases: Exploring the Dynamics of the S1′ Pocket in the Design of Selective, Small Molecule Inhibitors. *J. Med. Chem.* **2014**, *57* (24), 10205–10219. https://doi.org/10.1021/jm500505f.

(151) Rao, B. Recent Developments in the Design of Specific Matrix Metalloproteinase Inhibitors Aided by Structural and Computational Studies. *Curr. Pharm. Des.* **2005**, *11* (3), 295–322. https://doi.org/10.2174/1381612053382115.

(152) Nation, D. A.; Sweeney, M. D.; Montagne, A.; Sagare, A. P.; D'Orazio, L. M.; Pachicano, M.; Sepehrband, F.; Nelson, A. R.; Buennagel, D. P.; Harrington, M. G.; Benzinger, T. L. S.; Fagan, A. M.; Ringman, J. M.; Schneider, L; S; Morris, J. C.; Chui, H. C.; Law, M.; Toga, A. W.; Zlokovic, B. V. Blood–brain Barrier Breakdown Is an Early Biomarker of Human Cognitive Dysfunction. *Nat. Med.* **2019**, *25* (2), 270–276. https://doi.org/10.1038/s41591-018-0297-y.

(153) Montagne, A.; Zhao, Z.; Zlokovic, B. V. Alzheimer's Disease: A Matter of Blood–brain Barrier Dysfunction? *J. Exp. Med.* **2017**, *214* (11), 3151–3169. https://doi.org/10.1084/jem.20171406.

(154) Rempe, R. G.; Hartz, A. M. S.; Bauer, B. Matrix Metalloproteinases in the Brain and Blood–brain Barrier: Versatile Breakers and Makers. *J. Cereb. Blood Flow Metab.* **2016**, *36* (9), 1481–1507. https://doi.org/10.1177/0271678X16655551.

(155) Chaturvedi, M.; Kaczmarek, L. MMP-9 Inhibition: A Therapeutic Strategy in Ischemic Stroke. *Mol. Neurobiol.* **2014**, *49* (1), 563–573. https://doi.org/10.1007/s12035-013-8538-z.

(156) Al-Ahmady, Z. S. Selective Drug Delivery Approaches to Lesioned Brain through Blood Brain Barrier Disruption. *Expert Opin. Drug Deliv.* **2018**, *15* (4), 335–349. https://doi.org/10.1080/17425247.2018.1444601.

(157) Tam, V. H.; Sosa, C.; Liu, R.; Yao, N.; Priestley, R. D. Nanomedicine as a Non-Invasive Strategy for Drug Delivery across the Blood Brain Barrier. *Int. J. Pharm.* **2016**, *515* (1–2), 331–342. https://doi.org/10.1016/j.ijpharm.2016.10.031.

(158) Kumar, N. N.; Pizzo, M. E.; Nehra, G.; Wilken-Resman, B.; Boroumand, S.; Thorne, R. G. Passive Immunotherapies for Central Nervous System Disorders: Current Delivery Challenges and New Approaches. *Bioconjug. Chem.* **2018**, *29* (12), 3937–3966. https://doi.org/10.1021/acs.bioconjchem.8b00548.

(159) Barar, J.; Rafi, M. A.; Pourseif, M. M.; Omidi, Y. Blood-Brain Barrier Transport Machineries and Targeted Therapy of Brain Diseases. *BioImpacts* **2016**, *6* (4), 225–248. https://doi.org/10.15171/bi.2016.30.

(160) Gabathuler, R. Approaches to Transport Therapeutic Drugs across the Blood–brain Barrier to Treat Brain Diseases. *Neurobiol. Dis.* **2010**, *37* (1), 48–57. https://doi.org/10.1016/j.nbd.2009.07.028.

(161) Golub, L. M.; Lee, H.-M.; Ryan, M. E.; Giannobile, W. V.; Payne, J.; Sorsa, T. Tetracyclines Inhibit Connective Tissue Breakdown by Multiple Non-Antimicrobial

Mechanisms. *Adv. Dent. Res.* **1998**, *12* (1), 12–26. https://doi.org/10.1177/08959374980120010501.

(162) Blum, D.; Chtarto, A.; Tenenbaum, L.; Brotchi, J.; Levivier, M. Clinical Potential of Minocycline for Neurodegenerative Disorders. *Neurobiol. Dis.* **2004**, *17* (3), 359–366. https://doi.org/10.1016/j.nbd.2004.07.012.

(163) Teronen, O.; Heikkilä, P.; Konttinen, Y. T.; Laitinen, M.; Salo, T.; Hanemaaijer, R.; Teronen, A.; Maisi, P.; Sorsa, T. MMP Inhibition and Downregulation by Bisphosphonates. *Ann. N. Y. Acad. Sci.* **1999**, *878* (358), 453–465. https://doi.org/10.1111/j.1749-6632.1999.tb07702.x.

(164) Izidoro-Toledo, T. C.; Guimaraes, D. A.; Belo, V. A.; Gerlach, R. F.; Tanus-Santos, J. E. Effects of Statins on Matrix Metalloproteinases and Their Endogenous Inhibitors in Human Endothelial Cells. *Naunyn. Schmiedebergs. Arch. Pharmacol.* **2011**, *383* (6), 547–554. https://doi.org/10.1007/s00210-011-0623-0.

(165) Piura, B.; Medina, L.; Rabinovich, A.; Dyomin, V.; Huleihel, M. Thalidomide Distinctly Affected TNF-A, IL-6 and MMP Secretion by an Ovarian Cancer Cell Line (SKOV-3) and Primary Ovarian Cancer Cells. *Eur. Cytokine Netw.* **2013**, *24* (3), 122–129. https://doi.org/10.1684/ecn.2013.0342.

(166) Whittaker, M.; Floyd, C. D.; Brown, P.; Gearing, A. J. H. Design and Therapeutic Application of Matrix Metalloproteinase Inhibitors. *Chem. Rev.* **1999**, *99* (9), 2735–2776. https://doi.org/10.1021/cr9804543.

(167) Coussens, L. M. Matrix Metalloproteinase Inhibitors and Cancer: Trials and Tribulations.

*Science.* **2002**, *295* (5564), 2387–2392. https://doi.org/10.1126/science.1067100.

(168) Jacobsen, J. A.; Major Jourden, J. L.; Miller, M. T.; Cohen, S. M. To Bind Zinc or Not to Bind Zinc: An Examination of Innovative Approaches to Improved Metalloproteinase Inhibition. *Biochim. Biophys. Acta - Mol. Cell Res.* **2010**, *1803* (1), 72–94. https://doi.org/10.1016/j.bbamcr.2009.08.006.

(169) Jacobsen, J. A.; Fullagar, J. L.; Miller, M. T.; Cohen, S. M. Identifying Chelators for Metalloprotein Inhibitors Using a Fragment-Based Approach. *J. Med. Chem.* **2011**, *54* (2), 591–602. https://doi.org/10.1021/jm101266s.

(170) Agrawal, A.; Romero-Perez, D.; Jacobsen, J. A.; Villarreal, F. J.; Cohen, S. M. Zinc-Binding Groups Modulate Selective Inhibition of MMPs. *ChemMedChem* **2008**, *3* (5), 812–820. https://doi.org/10.1002/cmdc.200700290.

(171) Skiles, J.; Gonnella, N.; Jeng, A. The Design, Structure, and Clinical Update of Small Molecular Weight Matrix Metalloproteinase Inhibitors. *Curr. Med. Chem.* **2004**, *11* (22), 2911–2977. https://doi.org/10.2174/0929867043364018.

(172) Devel, L.; Czarny, B.; Beau, F.; Georgiadis, D.; Stura, E.; Dive Vincent, V. Third Generation of Matrix Metalloprotease Inhibitors: Gain in Selectivity by Targeting the Depth of the S1’ Cavity. *Biochimie* **2010**, *92* (11), 1501–1508. https://doi.org/10.1016/j.biochi.2010.07.017.

(173) Nuti, E.; Tuccinardi, T.; Rossello, A. Matrix Metalloproteinase Inhibitors: New Challenges in the Era of Post Broad-Spectrum Inhibitors. *Curr. Pharm. Des.* **2007**, *13* (20), 2087–2100. https://doi.org/10.2174/138161207781039706.

(174) Dormán, G.; Cseh, S.; Hajdú, I.; Barna, L.; Kónya, D.; Kupai, K.; Kovács, L.; Ferdinandy, P. Matrix Metalloproteinase Inhibitors. *Drugs* **2010**, *70* (8), 949–964. https://doi.org/10.2165/11318390-000000000-00000.

(175) Fisher, J. F.; Mobashery, S. Recent Advances in MMP Inhibitor Design. *Cancer Metastasis Rev.* **2006**, *25* (1), 115–136. https://doi.org/10.1007/s10555-006-7894-9.

(176) Tu, G.; Xu, W.; Huang, H.; Li, S. Progress in the Development of Matrix Metalloproteinase Inhibitors. *Curr. Med. Chem.* **2008**, *15* (14), 1388–1395. https://doi.org/10.2174/092986708784567680.

(177) Fischer, T.; Senn, N.; Riedl, R. Design and Structural Evolution of Matrix Metalloproteinase Inhibitors. *Chem. – A Eur. J.* **2019**, *25* (34), 7960–7980. https://doi.org/10.1002/chem.201805361.

(178) Jain, P.; Saravanan, C.; Singh, S. K. Sulphonamides: Deserving Class as MMP Inhibitors? *Eur. J. Med. Chem.* **2013**, *60*, 89–100. https://doi.org/10.1016/j.ejmech.2012.10.016.

(179) Nuti, E.; Cuffaro, D.; D'Andrea, F.; Rosalia, L.; Tepshi, L.; Fabbi, M.; Carbotti, G.; Ferrini, S.; Santamaria, S.; Camodeca, C.; Ciccone, L.; Orlandini, E.; Nencetti, S.; Stura, E. A.; Dive, V.; Rossello, A. Sugar-Based Arylsulfonamide Carboxylates as Selective and Water-Soluble Matrix Metalloproteinase-12 Inhibitors. *ChemMedChem* **2016**, 1626–1637. https://doi.org/10.1002/cmdc.201600235.

(180) Cuffaro, D.; Camodeca, C.; D'Andrea, F.; Piragine, E.; Testai, L.; Calderone, V.; Orlandini, E.; Nuti, E.; Rossello, A. Matrix Metalloproteinase-12 Inhibitors: Synthesis, Structure-Activity Relationships and Intestinal Absorption of Novel Sugar-Based

Biphenylsulfonamide Carboxylates. *Bioorganic Med. Chem.* **2018**, *26* (22), 5804–5815. https://doi.org/10.1016/j.bmc.2018.10.024.

(181) Nuti, E.; Cuffaro, D.; Bernardini, E.; Camodeca, C.; Panelli, L.; Chaves, S.; Ciccone, L.; Tepshi, L.; Vera, L.; Orlandini, E.; Nencetti, S.; Stura, E. A.; Santos, M. A. Development of Thioaryl-Based Matrix Metalloproteinase-12 Inhibitors with Alternative Zinc-Binding Groups: Synthesis, Potentiometric, NMR, and Crystallographic Studies. *J. Med. Chem.* **2018**, *61* (10), 4421–4435. https://doi.org/10.1021/acs.jmedchem.8b00096.

(182) Brown, S.; Bernardo, M. M.; Li, Z.; Kotra, L. P.; Tanaka, Y.; Fridman, R. Potent and Selective Mechanism-Based Inhibition of Gelatinases. *J. Am. Chem. Soc.* **2000**, *122* (12), 6799–6800. https://doi.org/10.1021/ja001461n.

(183) Kleifeld, O.; Kotra, L. P.; Gervasi, D. C.; Brown, S.; Bernardo, M. M.; Fridman, R.; Mobashery, S.; Sagi, I. X-Ray Absorption Studies of Human Matrix Metalloproteinase-2 (MMP-2) Bound to a Highly Selective Mechanism-Based Inhibitor. *J. Biol. Chem.* **2001**, *276* (20), 17125–17131. https://doi.org/10.1074/jbc.M011604200.

(184) Gooyit, M.; Suckow, M. A.; Schroeder, V. A.; Wolter, W. R.; Mobashery, S.; Chang, M. Selective Gelatinase Inhibitor Neuroprotective Agents Cross the Blood-Brain Barrier. *ACS Chem. Neurosci.* **2012**, *3* (10), 730–736. https://doi.org/10.1021/cn300062w.

(185) Ranasinghe, H. S.; Scheepens, A.; Sirimanne, E.; Mitchell, M. D.; Williams, C. E.; Fraser, M. Inhibition of MMP-9 Activity Following Hypoxic Ischemia in the Developing Brain Using a Highly Specific Inhibitor. *Dev. Neurosci.* **2012**, *34* (5), 417–427. https://doi.org/10.1159/000343257.

(186) Gooyit, M.; Song, W.; Mahasenan, K. V.; Lichtenwalter, K.; Suckow, M. A.; Schroeder, V. A.; Wolter, W. R.; Mobashery, S.; Chang, M. O -Phenyl Carbamate and Phenyl Urea Thiiranes as Selective Matrix Metalloproteinase-2 Inhibitors That Cross the Blood–Brain Barrier. *J. Med. Chem.* **2013**, *56* (20), 8139–8150. https://doi.org/10.1021/jm401217d.

(187) Jia, F.; Yin, Y. H.; Gao, G. Y.; Wang, Y.; Cen, L.; Jiang, J. MMP-9 Inhibitor SB-3CT Attenuates Behavioral Impairments and Hippocampal Loss after Traumatic Brain Injury in Rat. *J. Neurotrauma* **2014**, *31* (13), 1225–1234. https://doi.org/10.1089/neu.2013.3230.

(188) Devel, L.; Rogakos, V.; David, A.; Makaritis, A.; Beau, F.; Cuniasse, P.; Yiotakis, A.; Dive, V. Development of Selective Inhibitors and Substrate of Matrix Metalloproteinase-12. *J. Biol. Chem.* **2006**, *281* (16), 11152–11160. https://doi.org/10.1074/jbc.M600222200.

(189) Czarny, B.; Stura, E. A.; Devel, L.; Vera, L.; Cassar-Lajeunesse, E.; Beau, F.; Calderone, V.; Fragai, M.; Luchinat, C.; Dive, V. Molecular Determinants of a Selective Matrix Metalloprotease-12 Inhibitor: Insights from Crystallography and Thermodynamic Studies. *J. Med. Chem.* **2013**, *56* (3), 1149–1159. https://doi.org/10.1021/jm301574d.

(190) Rouanet-Mehouas, C.; Czarny, B.; Beau, F.; Cassar-Lajeunesse, E.; Stura, E. A.; Dive, V.; Devel, L. Zinc-Metalloproteinase Inhibitors: Evaluation of the Complex Role Played by the Zinc-Binding Group on Potency and Selectivity. *J. Med. Chem.* **2017**, *60* (1), 403–414. https://doi.org/10.1021/acs.jmedchem.6b01420.

(191) Devel, L.; Garcia, S.; Czarny, B.; Beau, F.; Lajeunesse, E.; Vera, L.; Georgiadis, D.; Stura, E.; Dive, V. Insights from Selective Non-Phosphinic Inhibitors of MMP-12

Tailored to Fit with an S1′ Loop Canonical Conformation. *J. Biol. Chem.* **2010**, *285* (46), 35900–35909. https://doi.org/10.1074/jbc.M110.139634.

(192) Devel, L.; Beau, F.; Amoura, M.; Vera, L.; Cassar-Lajeunesse, E.; Garcia, S.; Czarny, B.; Stura, E. A.; Dive, V. Simple Pseudo-Dipeptides with a P2′ Glutamate: A Novel Inhibitor Family of Matrix Metalloproteases and Other Metzincins. *J. Biol. Chem.* **2012**, *287* (32), 26647–26656. https://doi.org/10.1074/jbc.M112.380782.

(193) Nury, C.; Bregant, S.; Czarny, B.; Berthon, F.; Cassar-Lajeunesse, E.; Dive, V. Detection of Endogenous Matrix Metalloprotease-12 Active Form with a Novel Broad Spectrum Activity-Based Probe. *J. Biol. Chem.* **2013**, *288* (8), 5636–5644. https://doi.org/10.1074/jbc.M112.419499.

(194) Bertran, A.; Khomiak, D.; Konopka, A.; Rejmak, E.; Bulska, E.; Seco, J.; Kaczmarek, L.; Tarragó, T.; Prades, R. Design and Synthesis of Selective and Blood-Brain Barrier-Permeable Hydroxamate-Based Gelatinase Inhibitors. *Bioorg. Chem.* **2020**, *94* (10), 1–12. https://doi.org/10.1016/j.bioorg.2019.103365.

(195) Sela-Passwell, N.; Rosenblum, G.; Shoham, T.; Sagi, I. Structural and Functional Bases for Allosteric Control of MMP Activities: Can It Pave the Path for Selective Inhibition? *Biochim. Biophys. Acta - Mol. Cell Res.* **2010**, *1803* (1), 29–38. https://doi.org/10.1016/j.bbamcr.2009.04.010.

(196) Udi, Y.; Fragai, M.; Grossman, M.; Mitternacht, S.; Arad-Yellin, R.; Calderone, V.; Melikian, M.; Toccafondi, M.; Berezovsky, I. N.; Luchinat, C.; Sagi, I. Unraveling Hidden Regulatory Sites in Structurally Homologous Metalloproteases. *J. Mol. Biol.* **2013**,

*425* (13), 2330–2346. https://doi.org/10.1016/j.jmb.2013.04.009.

(197) Fields, G. B. New Strategies for Targeting Matrix Metalloproteinases. *Matrix Biol.* **2015**, *44–46*, 239–246. https://doi.org/10.1016/j.matbio.2015.01.002.

(198) Xie, X. W.; Wan, R. Z.; Liu, Z. P. Recent Research Advances in Selective Matrix Metalloproteinase-13 Inhibitors as Anti-Osteoarthritis Agents. *ChemMedChem* **2017**, *12* (15), 1157–1168. https://doi.org/10.1002/cmdc.201700349.

(199) Lauer-Fields, J. L.; Minond, D.; Chase, P. S.; Baillargeon, P. E.; Saldanha, S. A.; Stawikowska, R.; Hodder, P.; Fields, G. B. High Throughput Screening of Potentially Selective MMP-13 Exosite Inhibitors Utilizing a Triple-Helical FRET Substrate. *Bioorganic Med. Chem.* **2009**, *17* (3), 990–1005. https://doi.org/10.1016/j.bmc.2008.03.004.

(200) Roth, J.; Minond, D.; Darout, E.; Liu, Q.; Lauer, J.; Hodder, P.; Fields, G. B.; Roush, W. R. Identification of Novel, Exosite-Binding Matrix Metalloproteinase-13 Inhibitor Scaffolds. *Bioorg. Med. Chem. Lett.* **2011**, *21* (23), 7180–7184. https://doi.org/10.1016/j.bmcl.2011.09.077.

(201) Spicer, T. P.; Jiang, J.; Taylor, A. B.; Choi, J. Y.; Hart, P. J.; Roush, W. R.; Fields, G. B.; Hodder, P. S.; Minond, D. Characterization of Selective Exosite-Binding Inhibitors of Matrix Metalloproteinase 13 That Prevent Articular Cartilage Degradation in Vitro. *J. Med. Chem.* **2014**, *57* (22), 9598–9611. https://doi.org/10.1021/jm501284e.

(202) Choi, J. Y.; Fuerst, R.; Knapinska, A. M.; Taylor, A. B.; Smith, L.; Cao, X.; Hart, P. J.; Fields, G. B.; Roush, W. R. Structure-Based Design and Synthesis of Potent and Selective

Matrix Metalloproteinase 13 Inhibitors. *J. Med. Chem.* **2017**, *60* (13), 5816–5825. https://doi.org/10.1021/acs.jmedchem.7b00514.

(203) Nara, H.; Sato, K.; Naito, T.; Mototani, H.; Oki, H.; Yamamoto, Y.; Kuno, H.; Santou, T.; Kanzaki, N.; Terauchi, J.; Uchikawa, O.; Kori, M. Discovery of Novel, Highly Potent, and Selective Quinazoline-2-Carboxamide-Based Matrix Metalloproteinase (MMP)-13 Inhibitors without a Zinc Binding Group Using a Structure-Based Design Approach. *J. Med. Chem.* **2014**, *57* (21), 8886–8902. https://doi.org/10.1021/jm500981k.

(204) Nara, H.; Kaieda, A.; Sato, K.; Naito, T.; Mototani, H.; Oki, H.; Yamamoto, Y.; Kuno, H.; Santou, T.; Kanzaki, N.; Terauchi, J.; Uchikawa, O.; Kori, M. Discovery of Novel, Highly Potent, and Selective Matrix Metalloproteinase (MMP)-13 Inhibitors with a 1,2,4-Triazol-3-Yl Moiety as a Zinc Binding Group Using a Structure-Based Design Approach. *J. Med. Chem.* **2017**, *60* (2), 608–626. https://doi.org/10.1021/acs.jmedchem.6b01007.

(205) Dufour, A.; Sampson, N. S.; Li, J.; Kuscu, C.; Rizzo, R. C.; DeLeon, J. L.; Zhi, J.; Jaber, N.; Liu, E.; Zucker, S.; Cao, J. Small-Molecule Anticancer Compounds Selectively Target the Hemopexin Domain of Matrix Metalloproteinase-9. *Cancer Res.* **2011**, *71* (14), 4977–4988. https://doi.org/10.1158/0008-5472.CAN-10-4552.

(206) Remacle, A. G.; Golubkov, V. S.; Shiryaev, S. A.; Dahl, R.; Stebbins, J. L.; Chernov, A. V.; Cheltsov, A. V.; Pellecchia, M.; Strongin, A. Y. Novel MT1-MMP Small-Molecule Inhibitors Based on Insights into Hemopexin Domain Function in Tumor Growth. *Cancer Res.* **2012**, *72* (9), 2339–2349. https://doi.org/10.1158/0008-5472.CAN-11-4149.

(207) Scannevin, R. H.; Alexander, R.; Haarlander, T. M.; Burke, S. L.; Singer, M.; Huo, C.;

Zhang, Y. M.; Maguire, D.; Spurlino, J.; Deckman, I.; Carroll, K. I.; Lewandowski, F.; Devine, E.; Dzordzorme, K.; Tounge, B.; Milligan, C.; Bayoumy, S.; Williams, R.; Schalk-Hihi, C.; Leonard, K.; Jackson, P.; Todd, M.; Kuo, L. C.; Rhodes, K. J. Discovery of a Highly Selective Chemical Inhibitor of Matrix Metalloproteinase-9 (MMP-9) That Allosterically Inhibits Zymogen Activation. *J. Biol. Chem.* **2017**, *292* (43), 17963–17974. https://doi.org/10.1074/jbc.M117.806075.

(208) Alford, V. M.; Kamath, A.; Ren, X.; Kumar, K.; Gan, Q.; Awwa, M.; Tong, M.; Seeliger, M. A.; Cao, J.; Ojima, I.; Sampson, N. S. Targeting the Hemopexin-like Domain of Latent Matrix Metalloproteinase-9 (proMMP-9) with a Small Molecule Inhibitor Prevents the Formation of Focal Adhesion Junctions. *ACS Chem. Biol.* **2017**, *12* (11), 2788–2803. https://doi.org/10.1021/acschembio.7b00758.

(209) Mohan, V.; Talmi-Frank, D.; Arkadash, V.; Papo, N.; Sagi, I. Matrix Metalloproteinase Protein Inhibitors: Hightlighting a New Beginning for Metalloproteinases in Medicine. *Met. Med.* **2016**, *3*, 31–47. https://doi.org/10.2147/MNM.S119588.

(210) Fields, G. B. The Rebirth of Matrix Metalloproteinase Inhibitors: Moving Beyond the Dogma. *Cells* **2019**, *8* (9), 1–24. https://doi.org/10.3390/cells8090984.

(211) Li, K.; Tay, F. R.; Yiu, C. K. Y. The Past, Present and Future Perspectives of Matrix Metalloproteinase Inhibitors. *Pharmacol. Ther.* **2020**, *207*, 1–13. https://doi.org/10.1016/j.pharmthera.2019.107465.

(212) Santamaria, S.; de Groot, R. Monoclonal Antibodies against Metzincin Targets. *Br. J. Pharmacol.* **2019**, *176* (1), 52–66. https://doi.org/10.1111/bph.14186.

(213) Nam, D. H.; Rodriguez, C.; Remacle, A. G.; Strongin, A. Y.; Ge, X. Active-Site MMP-Selective Antibody Inhibitors Discovered from Convex Paratope Synthetic Libraries. *Proc. Natl. Acad. Sci.* **2016**, *113* (52), 14970–14975. https://doi.org/10.1073/pnas.1609375114.

(214) Lopez, T.; Mustafa, Z.; Chen, C.; Lee, K. B.; Ramirez, A.; Benitez, C.; Luo, X.; Ji, R.-R.; Ge, X. Functional Selection of Protease Inhibitory Antibodies. *Proc. Natl. Acad. Sci.* **2019**, *116* (33), 16314–16319. https://doi.org/10.1073/pnas.1903330116.

(215) Johnson, J. L.; Devel, L.; Czarny, B.; George, S. J.; Jackson, C. L.; Rogakos, V.; Beau, F.; Yiotakis, A.; Newby, A. C.; Dive, V. A Selective Matrix Metalloproteinase-12 Inhibitor Retards Atherosclerotic Plaque Development in Apolipoprotein E-Knockout Mice. *Arterioscler. Thromb. Vasc. Biol.* **2011**, *31* (3), 528–535. https://doi.org/10.1161/ATVBAHA.110.219147.

(216) Li, J. J.; Nahra, J.; Johnson, A. R.; Bunker, A.; O'Brien, P.; Yue, W.-S.; Ortwine, D. F.; Man, C.-F.; Baragi, V.; Kilgore, K.; Dyer, R. D.; Han, H-K.Quinazolinones and Pyrido[3,4- D ]Pyrimidin-4-Ones as Orally Active and Specific Matrix Metalloproteinase-13 Inhibitors for the Treatment of Osteoarthritis. *J. Med. Chem.* **2008**, *51* (4), 835–841. https://doi.org/10.1021/jm701274v.

(217) Settle, S.; Vickery, L.; Nemirovskiy, O.; Vidmar, T.; Bendele, A.; Messing, D.; Ruminski, P.; Schnute, M.; Sunyer, T. Cartilage Degradation Biomarkers Predict Efficacy of a Novel, Highly Selective Matrix Metalloproteinase 13 Inhibitor in a Dog Model of Osteoarthritis: Confirmation by Multivariate Analysis That Modulation of Type Ii Collagen and Aggrecan Degradation Pepti. *Arthritis Rheum.* **2010**, *62* (10), 3006–3015.

https://doi.org/10.1002/art.27596.

(218) Johnson, A. R.; Pavlovsky, A. G.; Ortwine, D. F.; Prior, F.; Man, C. F.; Bornemeier, D. A.; Banotai, C. A.; Mueller, W. T.; McConnell, P.; Yan, C.; Baragi, V.; Lesch, C.; Roark, W. H.; Wilson, M.; Datta, K.; Guzman, R.; Han, H. K.; Dyer, R. D. Discovery and Characterization of a Novel Inhibitor of Matrix Metalloprotease-13 That Reduces Cartilage Damage in Vivo without Joint Fibroplasia Side Effects. *J. Biol. Chem.* **2007**, *282* (38), 27781–27791. https://doi.org/10.1074/jbc.M703286200.

(219) Botkjaer, K. A.; Kwok, H. F.; Terp, M. G.; Karatt-Vellatt, A.; Santamaria, S.; McCafferty, J.; Andreasen, P. A.; Itoh, Y.; Ditzel, H. J.; Murphy, G. Development of a Specific Affinity-Matured Exosite Inhibitor to MT1-MMP That Efficiently Inhibits Tumor Cell Invasion in Vitro and Metastasis in Vivo. *Oncotarget* **2016**, *7* (13), 16773–16792. https://doi.org/10.18632/oncotarget.7780.

(220) Devy, L.; Huang, L.; Naa, L.; Yanamandra, N.; Pieters, H.; Frans, N.; Chang, E.; Tao, Q.; Vanhove, M.; Lejeune, A.; van Gool, R.; Sexton, D. J.; Kuang, G.; Rank, D.; Hogan, S.; Pazmany, C.; Ma, Y. L.; Schoonbroodt, S.; Nixon, A. E.; Ladner, R. C.; Hoet, R.; Henderikx, P.; TenHoor, C.; Rabbani, S. A.; Valentino, M. L.; Wood, C. R.; Dransfield, D. T. Selective Inhibition of Matrix Metalloproteinase-14 Blocks Tumor Growth, Invasion, and Angiogenesis. *Cancer Res.* **2009**, *69* (4), 1517–1526. https://doi.org/10.1158/0008-5472.CAN-08-3255.

(221) Ager, E. I.; Kozin, S. V.; Kirkpatrick, N. D.; Seano, G.; Kodack, D. P.; Askoxylakis, V.; Huang, Y.; Goel, S.; Snuderl, M.; Muzikansky, A.; Finkelstein, D. M.; Dransfield, D. T.; Devy, L.; Boucher, Y.; Fukumura, D.; Jain, R. K. Blockade of MMP14 Activity in Murine

Breast Carcinomas: Implications for Macrophages, Vessels, and Radiotherapy. *JNCI J. Natl. Cancer Inst.* **2015**, *107* (4), 1–12. https://doi.org/10.1093/jnci/djv017.

(222) Kaneko, K.; Williams, R. O.; Dransfield, D. T.; Nixon, A. E.; Sandison, A.; Itoh, Y. Selective Inhibition of Membrane Type 1 Matrix Metalloproteinase Abrogates Progression of Experimental Inflammatory Arthritis: Synergy With Tumor Necrosis Factor Blockade. *Arthritis Rheumatol.* **2016**, *68* (2), 521–531. https://doi.org/10.1002/art.39414.

(223) Fischer, T.; Riedl, R. Development of a Non-Hydroxamate Dual Matrix Metalloproteinase (MMP)-7/-13 Inhibitor. *Molecules* **2017**, *22* (9), 1–27. https://doi.org/10.3390/molecules22091548.

(224) Gall, F. M.; Hohl, D.; Frasson, D.; Wermelinger, T.; Mittl, P. R. E.; Sievers, M.; Riedl, R. Drug Design Inspired by Nature: Crystallographic Detection of an Auto-Tailored Protease Inhibitor Template. *Angew. Chemie Int. Ed.* **2019**, *58* (12), 4051–4055. https://doi.org/10.1002/anie.201812348.

(225) El Ashry, E. S. H.; Awad, L. F.; Teleb, M.; Ibrahim, N. A.; Abu-Serie, M. M.; Abd Al Moaty, M. N. Structure-Based Design and Optimization of Pyrimidine- and 1,2,4-triazolo[4,3-A]pyrimidine-Based Matrix Metalloproteinase-10/13 Inhibitors via Dimroth Rearrangement towards Targeted Polypharmacology. *Bioorg. Chem.* **2020**, *96* (1), 1–17. https://doi.org/10.1016/j.bioorg.2020.103616.

(226) Swetha, R.; Kumar, D.; Gupta, S. K.; Ganeshpurkar, A.; Singh, R.; Gutti, G.; Kumar, D.; Jana, S.; Krishnamurthy, S.; Singh, S. K. Multifunctional Hybrid Sulfonamides as Novel

Therapeutic Agents for Alzheimer's Disease. *Future Med. Chem.* **2019**, *11* (24), 3161–3178. https://doi.org/10.4155/fmc-2019-0106.

**Table of Contents graphic**

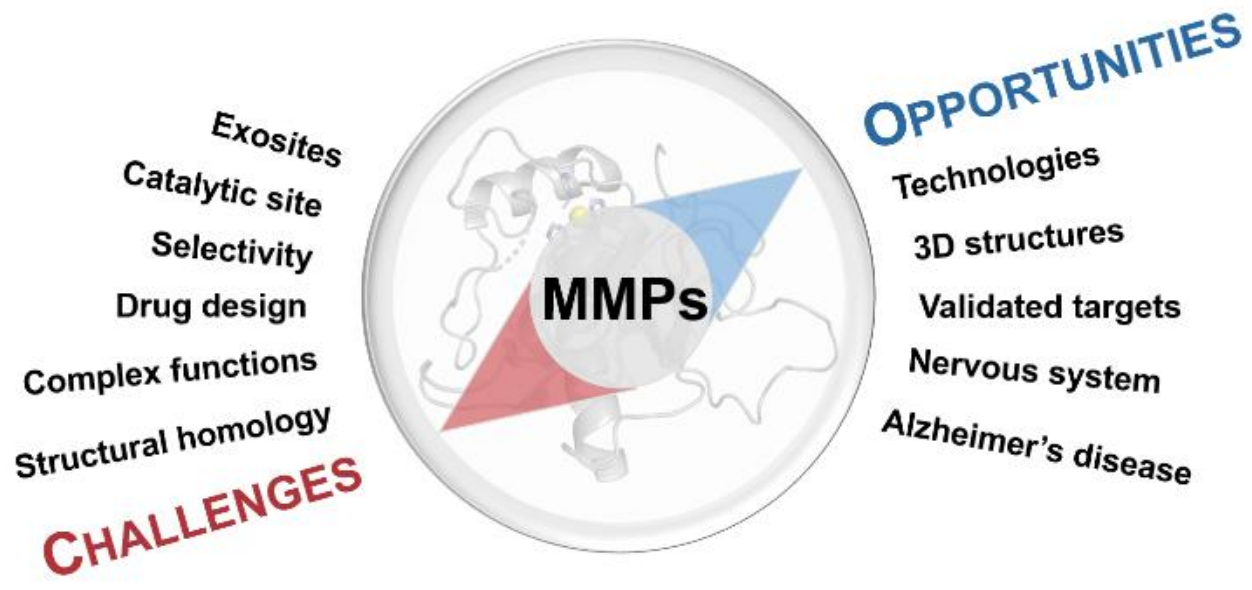